\documentclass[structabstract]{aa}  
\usepackage{natbib}  
\bibpunct{(}{)}{;}{a}{}{,} 
\usepackage{graphicx}   

\usepackage{txfonts} 
\usepackage{lscape}   
\renewcommand\footnoterule{  \kern -3pt  \hrule \kern 2pt }

\begin{document}
\title{Integral field spectroscopy of M1-67. A Wolf-Rayet nebula with LBVN appearance. \thanks{Based on observations collected at the Centro Astron\'omico Hispano Alem\'an (CAHA) at Calar Alto, operated jointly by the Max-Planck-Institut f\"ur Astronomie and the Instituto de Astrof\'isica de Andaluc\'ia (CSIC).}}

\author{A. Fern\'andez-Mart\'in \inst{1} \thanks{e-mail: alba@iaa.es} ,
							 J.M. V\'ilchez \inst{1} , 
							E. P\'erez-Montero  \inst{1},
							A. Candian \inst{1,2},
                			             S. F. S\'anchez  \inst{1,3},
							D. Mart\'in-Gord\'on \inst{1},
							 \and 
						     A. Riera \inst{4} 
				}
\institute{Instituto de Astrof\'isica de Andaluc\'ia (IAA-CSIC),  Glorieta de la Astronom\'ia  S/N, 18008 Granada, Spain 
						    \and Leiden Observatory, Niels Bohrweg 2, 2333 CA Leiden, The Netherlands
						    \and Centro Astron\'omico Hispano Alem\'an, Calar Alto, CSIC-MPG, C/Jesús Durb\'an Remón 2-2, E-04004 Almer\'ia, Spain
						   \and Departament de F\'isica i Enginyeria Nuclear, EUETIB, Universitat Polit\'ecnica de Catalunya, C. Comte Urgell 187, 08036, Barcelona, Spain
				}                                                                       



\abstract 
{}
{This work aims to disentangle the morphological, kinematic, and chemical components of the nebula M1-67 to shed light on its process of formation around the central Wolf-Rayet (WR) star WR124.}
{We have carried out integral field spectroscopy observations over two regions of M1-67, covering most of the nebula in the optical range. Maps of electron density, line ratios, and radial velocity were created to perform a detailed analysis of the two-dimensional structure. We studied the physical and chemical properties by means of integrated spectra selected over the whole nebula. Photoionization models were performed to confirm the empirical chemical results theoretically. In addition, we obtained and analysed infrared spectroscopic data and the MIPS 24$\mu$m image of M1-67 from Spitzer.}
{We find that the ionized gas of M1-67 is condensed in knots aligned in a preferred axis along the NE-SW direction, like a bipolar structure. Both electron density and radial velocity decrease in this direction when moving away from the central star. From the derived electron temperature, T$_{e}\sim$8200~K, we have estimated chemical abundances, obtaining that nitrogen appears strongly enriched and oxygen depleted. From the last two results, we infer that this bipolarity is the consequence of an ejection of an evolved stage of WR124 with material processed in the CNO cycle. Furthermore, we find two regions placed outside of the bipolar structure with different spectral and chemical properties. The infrared study has revealed that the bipolar axis is composed of ionized gas with a low ionization degree that is well mixed with warm dust and of a spherical bubble surrounding the ejection at 24$\mu$m.

Taking the evolution of a 60~M$\sun$ star and the temporal scale of the bipolar ejection into account , we propose that the observed gas was ejected during an eruption in the luminous blue variable. The star has entered the WR phase recently without apparent signs of interaction between WR-winds and interstellar material.}
{}


\titlerunning{Integral Field Spectroscopy Study of M1-67}
\authorrunning{Fern\'andez-Mart\'in et al.}
\maketitle

\section{Introduction \label{intro}}
WR124 ($\equiv$BAC209) is a Galactic massive star characterized by a very high heliocentric recession velocity of $\sim$175 km~s$^{-1}$ \citep{1982A&A...116...54S}, and it  is regarded to be among the fastest moving massive stars in the Galaxy \citep{1982A&A...114..135M}. It was classified by \citet{1938PASP...50..350M} as a nitrogen-sequence Wolf-Rayet star (WN) and later as Population I WN8 star \citep{1969ApJ...157.1245S}. \\ 
 
WR stars are thought to be a late stage in the evolution of stars more massive than 25~$M_{\mathrm{\sun}}$ and they are characterized by significant stellar winds with high mass-loss rate and terminal velocity. Many WRs are surrounded by nebular emission, some of which are members of a class of object called \textit{ring nebulae}. The structure of this type of nebula is attributed to a continual process of mass loss from the exciting WR star, which sweeps the surrounding interestellar gas into a shell \citep{1965ApJ...142.1033J}. The study of nebulae around WRs gives us clues to the mass-loss history of massive stars, as well as to the chemical enrichment of the interstellar medium (ISM).\\

\begin{table*}[t]
		 \caption{Main physical parameters of WR\,124 and M1-67.}   
		\label{table:parameter}    
		\centering                       
        	\begin{tabular}{ l l l l l}	\hline
				\\
             Object &  Parameter   &  Value   & Reference   \\  \hline \hline \\
			WR\,124 & ($\alpha$,$\delta$) (J2000) & (19:11:30.88, +16:51:38.16) & \citet{1997ESASP1200.....P} \\ 	
			& Spectral type & WN 8 &  \citet{1969ApJ...157.1245S} \\
		    & $v_{\mathrm{\infty}}$  (km~s$^{-1}$)  & 710 &   \citet{2001NewAR..45..135V} \\
		    & $T_{\mathrm{eff}}$  (kK)  & 44.7 & Hamann et al. (2006) \\
			&  Distance (kpc )& 4-5  & Crawford \& Barlow (1991)  \\
			& $R_{\mathrm{G}} $ (kpc )& 8-10&  Esteban et al. (1992)   \\
			&	 $v_{\mathrm{hel}}$  (km~s$^{-1}$)  &  175 &  Solf \& Carsenty (1982)  \\
     		& $M_{\mathrm{v}} $ (mag) & -7.22 &  Hamann et al. (2006)  \\
            & $E_{\mathrm{b-v}} $ (mag) & 1.08  &  Hamann et al. (2006) \\ 	 		
     \\ 
			M1-67 & H${\alpha}$ diameter (arcsec) & 110-120 & \citet{1998ApJ...506L.127G}   \\
			&	 $v_{\mathrm{hel}}$  (km~s$^{-1}$)  & 150-185  &   \citet{1981ApJ...249..586C} \\
			&	 $v_{\mathrm{exp}}$  (km~s$^{-1}$)  &  46 &  Sirianni et al. (1998) \\
			&$ M_{\mathrm{ionized}}$ ($M_{\mathrm{\sun}}$) &  1.73  &   \citet{1998ApJ...506L.127G} \\
				\hline		\\ 
			\end{tabular}
		\end{table*}

M1-67 \citep[$\equiv$Sh2-80,][]{1959ApJS....4..257S} is a bright nebula surrounding WR124, and it shows a clumpy and irregular distribution of gas that is mostly condensed in bright knots and filaments. It was first reported by \citet{1946PASP...58..305M} during an H${\alpha}$ objective prism survey. Classification of the nebula and its distance have been subjects of debate in the past years. Although it was first considered an H{\sc ii} region \citep{1959ApJS....4..257S}, the classification of M1-67 has been alternating between a planetary nebula (PN) and a ring nebula. \citet{1964PASP...76..241B} adopted a distance of 0.9~kpc and suggested that M1-67 might be a PN since both star and nebula have high radial velocity. Studies from optical, infrared, and radio data by \citet{1975ApL....16..165C} prompted its classification as ring nebula around a WR, with a distance of 4.33~kpc (in agreement with \citet{1979RMxAA...4..271P} estimations). Nevertheless, \citet{1985A&A...145L..13V} supported the PN status based on the energy distribution in the far infrared. The issue was settled by  \citet{1991A&A...244..205E} and \citet{1991A&A...249..518C}. The detailed abundance analysis of the nebula by \citet{1991A&A...244..205E} revealed nitrogen enhancement and oxygen deficiency, which is typical of material ejected in a previous evolutionary phase, and pointed to a progenitor more massive than those usually associated with PN central stars. \citet{1991A&A...249..518C} estimated a distance between 4 kpc and 5 kpc using the interstellar Na{\sc i}D$_{2}$ absorption spectrum of the star, ruling out the PN nature. Recently, \citet{2010ApJ...724L..90M} have used a comprehensive model of the nebular expansion to estimate a distance of 3.35~kpc. Currently, M1-67 is classified as an ejected type WR ring-nebula.

Although M1-67 shows an apparent spherical symmetry, ground-based coronographic images revealed a bipolar structure \citep{1995IAUS..163...78N}. The emission lines seems to be caused by condensations of gas in clumps and radial filaments \citep{1998ApJ...506L.127G}. One of the most striking characteristics of the nebula is the virtual absence of optical oxygen emission lines \citep{1978ApJ...219..914B,1991A&A...244..205E}. Nevertheless, \citet{1981ApJ...249..586C} reported a bright spot of [O{\sc iii}]$\lambda$5007\AA{} 15\arcsec\, to the NE of the central star. Spectroscopic investigations of the physical conditions and abundances of the nebular shell have shown that the ionized gas is nitrogen-enriched and oxygen-depleted, suggesting that O has been processed into N mainly via the ON cycle \citep{1991A&A...244..205E}. This implies that M1-67 is almost completely composed of stellar material that is poorly mixed with the surrounding ISM. The long-slit spectroscopy of M1-67 established that the bulk of the nebula is expanding at $v_{\mathrm{exp}}$=42-46~km~s$^{-1}$ \citep{1982A&A...116...54S,1998A&A...335.1029S} and \citet{1981ApJ...249..586C} confirmed the high heliocentric velocity of the nebula $v_{\mathrm{hel}}$=150-185~km~s$^{-1}$, which is comparable to the velocity of the star. The main parameters of the central star, WR124, and the nebula, M1-67, are summarized in Table \ref{table:parameter}. 

Many studies have tried to disentangle the geometry and dynamics of M1-67 and its interaction with WR124. \citet{1982A&A...116...54S} proposed a simple expanding \textquotedblleft empty\textquotedblright shell with condensation of stellar material that was ejected by the high-velocity parent star; indeed, the leading edge of the shell is considerably brighter than the trailing part. \citet{1998A&A...335.1029S} found two components in the environment of the central star and interpreted them as the consequence of two different events in the past: a spherical hollow shell of 92\arcsec\, in diameter expanding at 46~km~s$^{-1}$ and a bipolar outflow with a semi-dimension of 48\arcsec\, and a velocity of 88~km~s$^{-1}$ with respect to the expansion centre. On the other hand, some authors explained the asymmetry as the result of a possible low-mass companion for WR124 \citep{1981ApJ...249..586C,1982A&A...114..135M}. \\

Despite the important findings of the last years, some relevant aspects of the evolution and formation of the ring nebula associated with WR124 remain unknown. In particular, a 2D study of the ionization structure of the nebula covering all the morphologies and/or the structural components can shed light on the formation process of the nebula from the ejecta of the central star. The late spectral type of the WR ionizing star (WN8) is also very remarkable to study, as well as the degree of homogeneity in the chemical composition of its ejecta. \\

To do this, we included M1-67 in our programme of integral field spectroscopy (IFS) observations to compare the 2D structure with the integrated properties of certain selected areas and with models of WR evolution. The paper is organized as follows. First, we describe the observations and data reduction in Sect. \ref{obsandred}. Then, we present the 2D results for morphology, ionization structure, and kinematic in Sect. \ref{2d}. In Sect. \ref{1d} we show the physical conditions and chemical abundances of eight selected areas. We perform a study of M1-67 in the mid-infrared range by analysing the IRS spectrum and the 24$\,\mu$m MIPS image from Spitzer in Sect. \ref{ir}. In Sect. \ref{discussion} we discuss the chemical composition of M1-67, the observed structure, and its relation with the evolution of the central WR star. Finally, a summary of the main conclusion is given in Sect. \ref{conclusions}.

\section{Observation and data reduction \label{obsandred}} 
\begin{figure*}
\centering
\includegraphics[width=14cm]{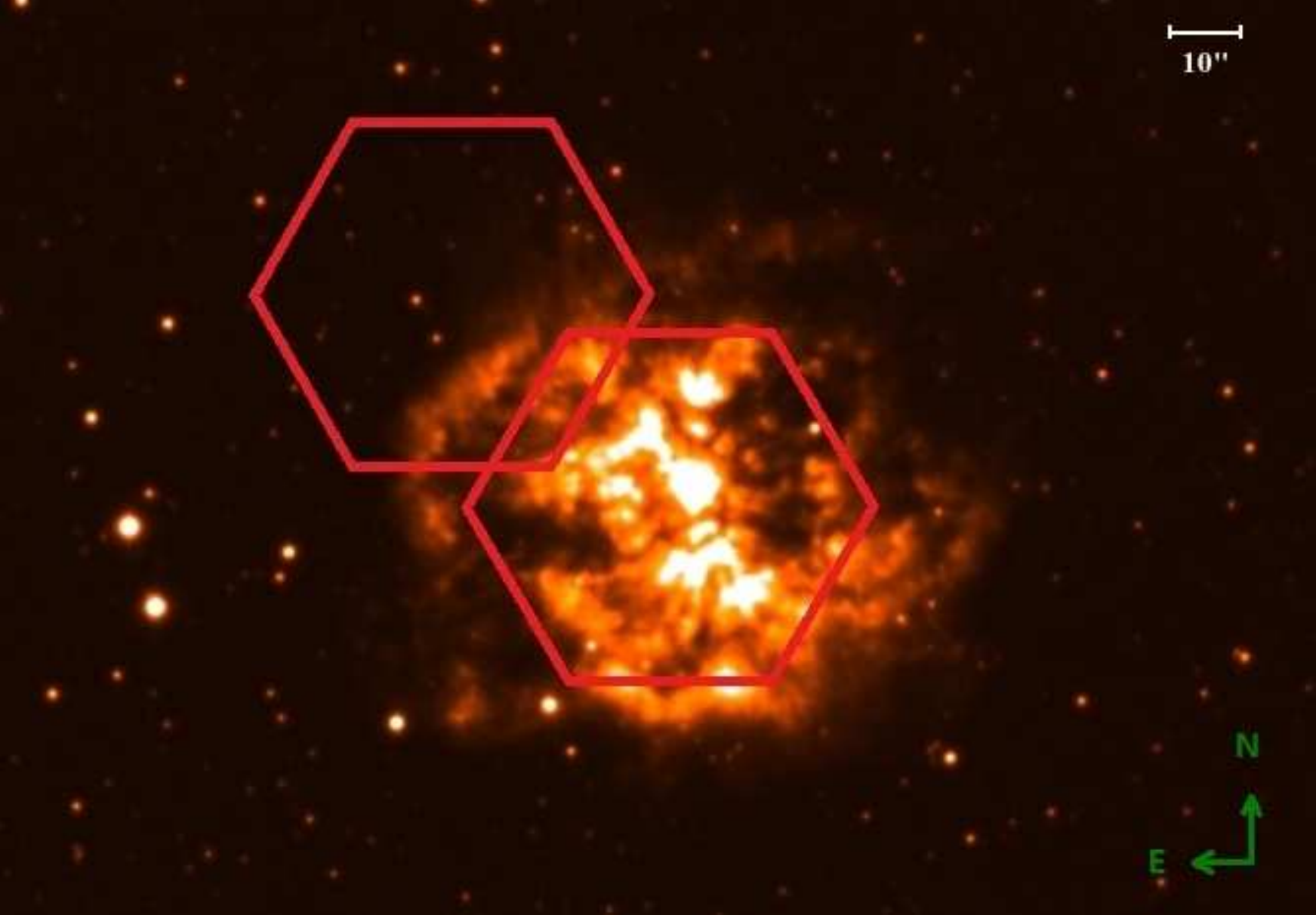}
\caption{Narrow-band image of M1-67 in H${\alpha}$+continuum taken with the Wide Field Camera at the Isaac Newton Telescope. North is up and east left. Red hexagons show the two zones of our IFS observations: \emph{Edge} to the NE (left) and \emph{Centre} to the SW (right).}
\label{fig:rgb}
\end{figure*}

The observations were carried out on July 5, 2005 using the Potsdam Multi-Aperture Spectrograph instrument (PMAS) \citep{2005PASP..117..620R} in PPAK mode (PMAS fibre Package, \citealt{2006PASP..118..129K}) at the 3.5~m telescope of the Centro Astron\'omico Hispano Alem\'an (CAHA) at the observatory of Calar Alto (Almer\'ia, Spain). 
The PPAK fibre bundle consists of 382 fibres with a diameter of 2.7 arcsec. The 331 science fibres are concentrated in an hexagonal bundle covering a field of view (FoV) of $74\arcsec \times 65\arcsec$. The surrounding sky is sampled by 36 fibres distributed in six bundles located following a circle at about 90\arcsec \,from the centre. There are 15 fibres for calibration purposes too \citep[see Fig. 5 in][]{2006PASP..118..129K}. We used the V300 grating, covering the spectral range from 3660 to 7040 \AA{} with a dispersion of 1.67~\AA{}/pix, giving a spectral resolution of FWHM$\sim$8.7 \AA{} (R = $\lambda/\Delta\lambda\sim$ 660) at 5577\AA{}. The weather was photometric throughout the observations with the typical seeing subarsecond. \\

To choose the regions of M1-67 to be mapped, we resorted to the narrow-band images observed by our group at the Isaac Newton Telescope (INT) with the Wide Field Camera (WFC). The first PPAK pointing (called \emph{Centre}) was centred on the WR star and it covers almost the whole nebula. The second zone (called \emph{Edge}) was selected to study the NE edge of the object containing nebular emission and surrounding medium. Both regions can be seen in Fig. \ref{fig:rgb}. Table \ref{table:log} shows the observational log for M1-67.

Bias frames, continuum, arcs, and one spectrophotometric standard star (Hz\,44) were also acquired during the observations.\\

\begin{table*}
\caption{M1-67 PPAK observational log.} 
\label{table:log} 
\centering 
\begin{tabular}{l c c c c c c }
\hline
Zone & Coordinates (J2000) &Grating & Spectral range & Exp. time & Airmass & Date \\
& ($\alpha$,$\delta$) & & ( \AA{} ) & (s) & &\\
\hline\hline
\\
Centre & (19:11:30.9 , +16:51:39.2) & V300 & 3640-7040 & 3 $\times$ 30 & 1.08 & July, 5, 2005 \\
Edge & (19:12:14.8 , +16:52:12.9) & V300 & 3640-7040 & 3 $\times$ 450 & 1.07 &July, 5, 2005 \\
\\
\hline
\end{tabular}
\end{table*}

The data were reduced using the R3D software \citep{2006AN....327..850S} in combination with IRAF\footnote{The Image Reduction and Analysis Facility IRAF is distributed by the National Optical Astronomy Observatories, which are operated by Association of Universities for Research in Astronomy, Inc., under cooperative agreement with the National Science Foundation. Website: http://iraf.noao.edu/.} and the Euro3D packages \citep{2004AN....325..167S}. The reduction consisted of the standard steps for fibre-fed IFS observations.

At first, a master bias frame was created and subtracted from all the images. The different exposures taken at the same position on the sky were combined to reject cosmic rays using IRAF routines. A trace mask was generated from a continuum-illuminated exposure, identifying the location of each spectrum on the detector along the dispersion axis. Then, each spectrum was extracted from the science and standard star frames, co-adding the flux within an aperture of five pixels at the location of the spectral peak in the raw data using the tracing information, and storing it in a 2D image called row-stacked spectrum (RSS) \citep{2004AN....325..167S}. We checked that the contamination from flux coming from adjacent fibres using this aperture was negligible \citep{2004PASP..116..565B,2006AN....327..850S}. For a given aperture and $FWHM\sim(0.5\times~aperture)$, we found a level of cross-talk that was always $<$10$\%$. This seems to be an acceptable compromise between maximizing the recovered flux and minimizing the cross-talk. 

Distortion and dispersion solutions were obtained using a He calibration-lamp exposure and applied to the science data to perform the wavelength calibration. The accuracy achieved was better than $\sim$0.1~\AA{} (rms) for the arc exposures. Corrections to minimize the differences between fibre-to-fibre transmission throughput were also applied, creating a fibre flat from the exposure of a continuum source. Observations of the spectrophotometric standard star Hz\,44 were used to perform the flux calibration. 

The sky emission was determined using the science data spectra obtained throughout the additional fibres for sampling the sky. As explained above, the second pointing was made at the edge of M1-67, and some of its sky bundles are located within an area containing signals from the nebula. We inspected the 36 sky-fibres of each pointing, selecting those that did not show nebular emission. The spectra of all the selected fibres were combined with a mean in a single spectrum, and a 2D spectrum was created by copying the combined spectrum in each fibre. These sky spectra were then subtracted from every science spectrum, each pointing with its own sky.

Finally, considering the wavelength range and the airmass of the observations, and using \citet{1982PASP...94..715F}, we estimated that offsets due to the differential atmospheric refraction (DAR) were always smaller than one third of the fibre diameter. Correction for DAR was not necessary in our data.

\section{Two-dimensional analysis \label{2d}} 
To perform a detailed analysis of the 2D structure of the nebula, we built easy-to-use datacubes of the two reduced pointings with three dimensions: two spatials and one spectral. We studied all the interpolation routines included in the E3D software and verified which conserves the flux and the apparent morphology observed in the spaxels. Finally, we generated our cubes with the linear Delaunay interpolation and a pixel size of 1.5$\times$1.5~arcsec$^{2}$.\\

\begin{figure}
\includegraphics[width=9cm]{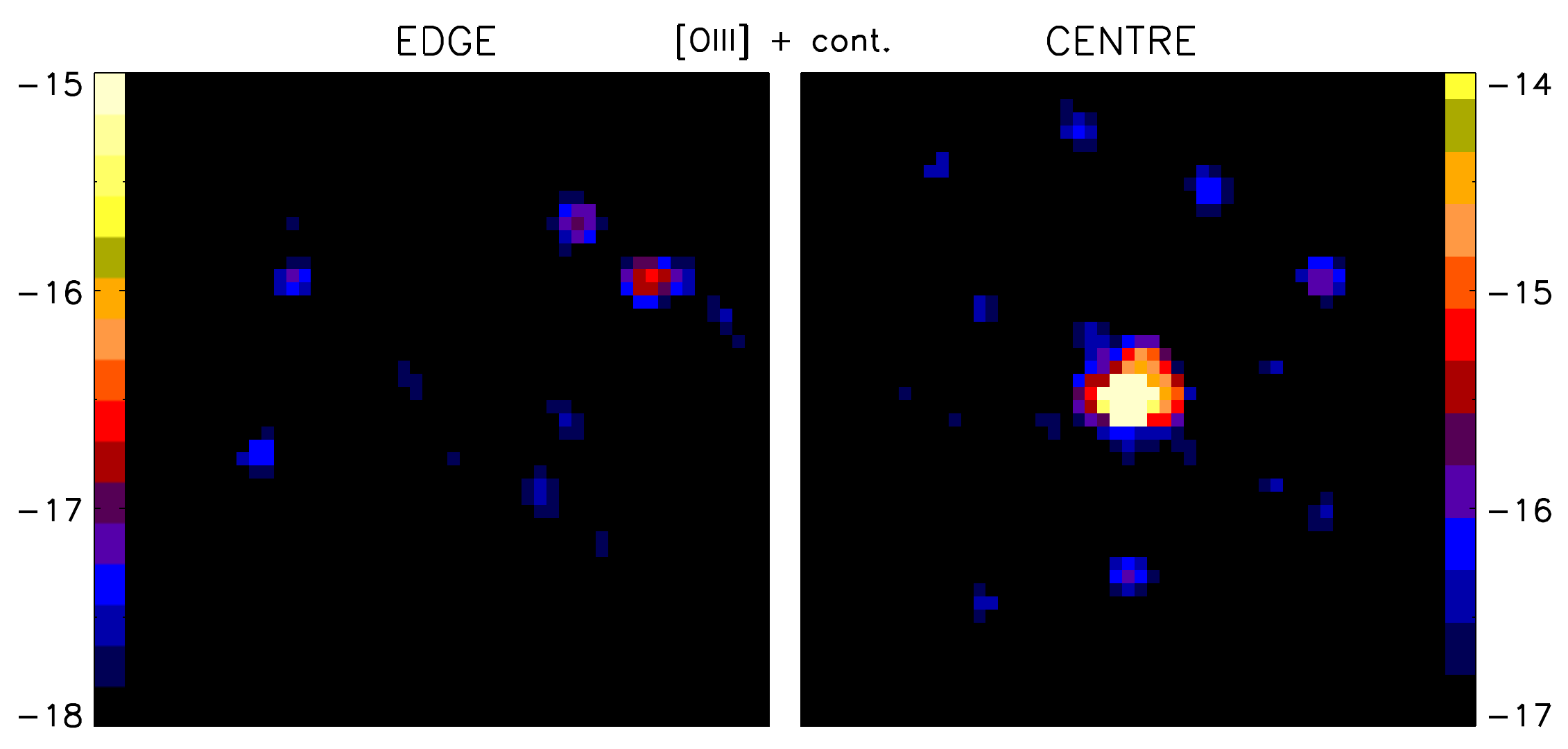}
\includegraphics[width=9cm]{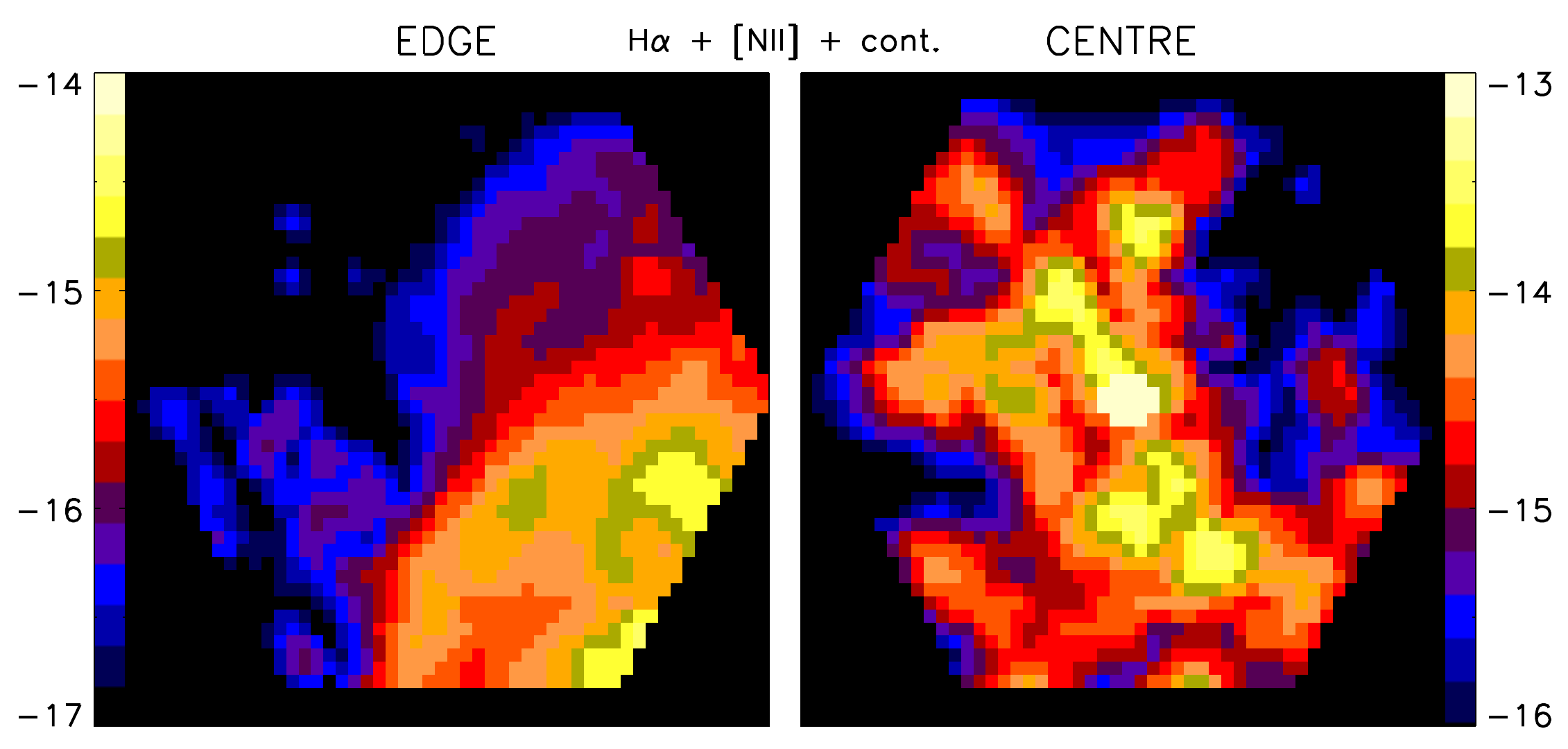}
\includegraphics[width=9cm]{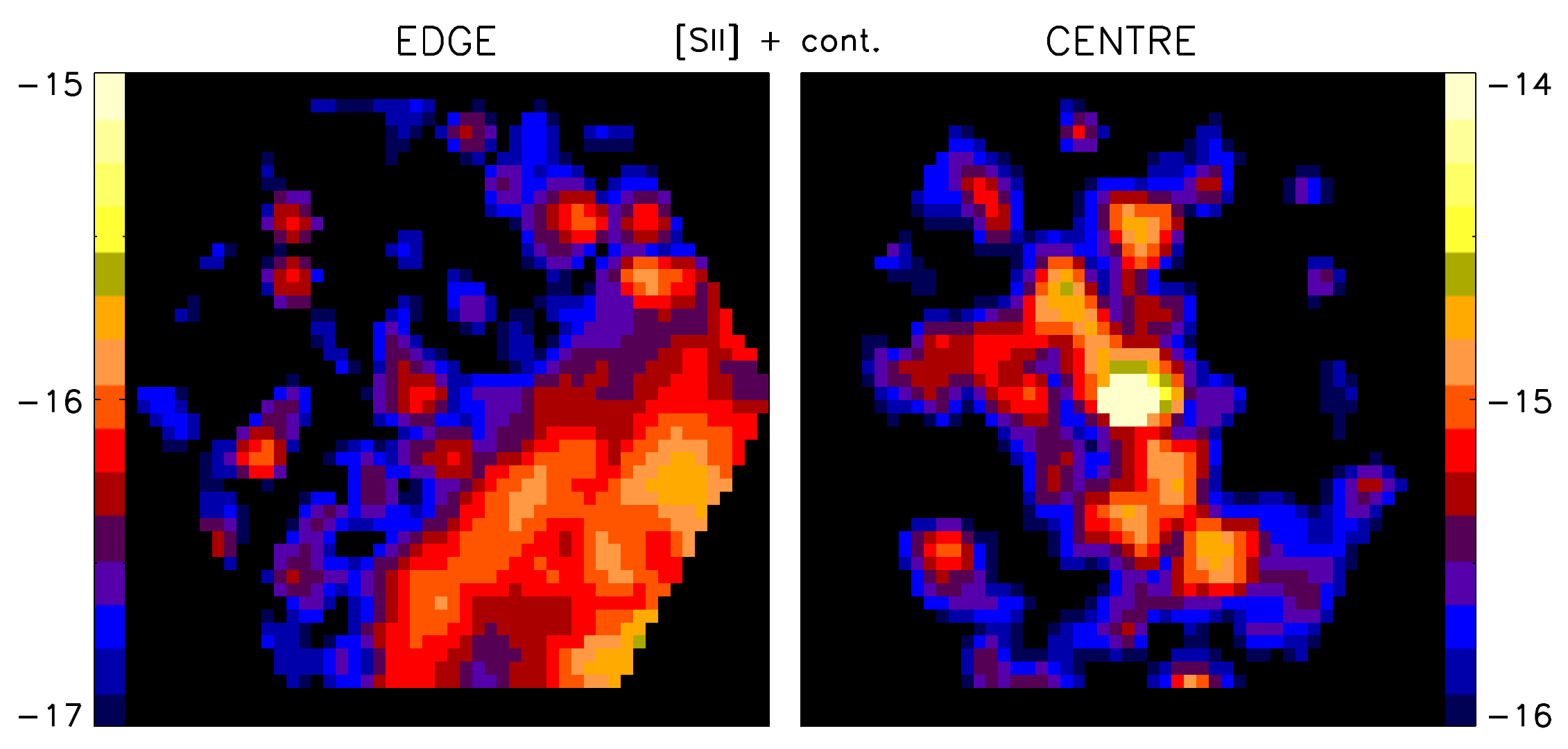}
\caption{Interpolated images of M1-67 of the two observed regions: In the left column the edge pointing and in the right the central one. In each row we represent the flux (including continuum) integrated in a wavelength range. Top: range 5006\AA{}-5014\AA{} including [O{\sc iii}]$\lambda$5007\AA{}. Middle: range 6562\AA{}-6590\AA{} including H${\alpha}$ and [N{\sc ii}]$\lambda$6584\AA{}. Bottom: range 6729\AA{}-6737\AA{} including [S{\sc ii}]$\lambda \lambda$6717,6731$\AA{}$. 
All the maps are represented on logarithmic scales with units of $\log$(erg~cm$^{-2}$~s$^{-1}$). The size of the hexagon side is 38\arcsec. In all the maps, north is up and east to the left (see Fig. \ref{fig:rgb}).}
\label{fig:morphology_all}
\end{figure}

In a first attempt to understand the morphology of the two observed zones, we extracted images by sliding the cubes at different wavelength ranges. They are presented in Fig. \ref{fig:morphology_all} on logarithmic scale.

In the 5006\AA{}-5014\AA{} range, which includes the [O{\sc iii}]$\lambda$5007\AA{} line, no significant extended emission can be observed in both regions, supporting previous studies that revealed no oxygen emission in M1-67. Several spots appear in the FoV (including the central WR124 star) with fluxes lower than
$\sim10^{-17}$~erg~cm$^{-2}$~s$^{-1}$, but we checked that their nature was not nebular. They probably are stars in our line of sight. We gave special attention to the spot described by \citet{1981ApJ...249..586C} at 15\arcsec\ to the NE of the star. Although we can observe some emission, we cannot confirm that it comes from the nebula. A more detailed analysis of this spot is performed in Sect. \ref{1d} by means of the integrated spectrum. 

As for other lines from the central pointing maps (H${\alpha}$, [N{\sc ii}], and [S{\sc ii}]), most of the emission seems to be concentrated in at least five knots distributed in the NE-SW direction without counterpart in the [O{\sc iii}]$\lambda$5007\AA{} image. In addition, two regions with very faint surface brightness (or even no emission) can be seen at the opposite sides (NW and SE). This orientation agrees with the bipolar structure observed by \citet{1995IAUS..163...78N} and \citet{1998A&A...335.1029S} with coronographic studies. H${\alpha}$ and [S{\sc ii}] emission shows a discontinuity in the edge pointing of the nebula with higher surface brightness in its SW area. When we move to the NE, the emission decreases until it disappears. The purple coloured area reaches non-negligible emission up to H${\alpha}\sim10^{-16}$~erg~cm$^{-2}$~s$^{-1}$ per pixel (1pixel=2.25~arcsec$^2$).\\

\subsection{2D study of the emission-line maps \label{maps}}
Maps were created from cubes by fitting the emission lines in each spatial element following the methodology presented in \citet{2012A&A...541A.119F}. Basically, we performed a Gaussian fit to the emission lines using our own routine, which returns maps of the flux, centre, and FWHM among other properties. To prevent contamination by low signal-to-noise (S/N) data, we masked out all pixels with S/N lower than 5. During the creation of the S/N masks, we visually inspected all the pixels, rejecting interactively those with non-nebular spectra or with contamination from the central WR and other field stars.

For both regions (centre and edge pointings), maps of parameters from the Gaussian fitting were generated for seven emission lines: H${\gamma}$, H${\beta}$, H${\alpha}$, [N{\sc ii}]$\lambda \lambda$6548,6584\AA{}, and [S{\sc ii}]$\lambda \lambda$6717,6731$\AA{}$. Although our spectral range includes the [O{\sc ii}]$\lambda \lambda$3726,3728\AA{} lines, their automatic fit was not considered because these lines are faint and placed at the edge of the CCD, where the distortion correction bended and deformed them.\\

\begin{figure}
\centering
\includegraphics[width=9cm]{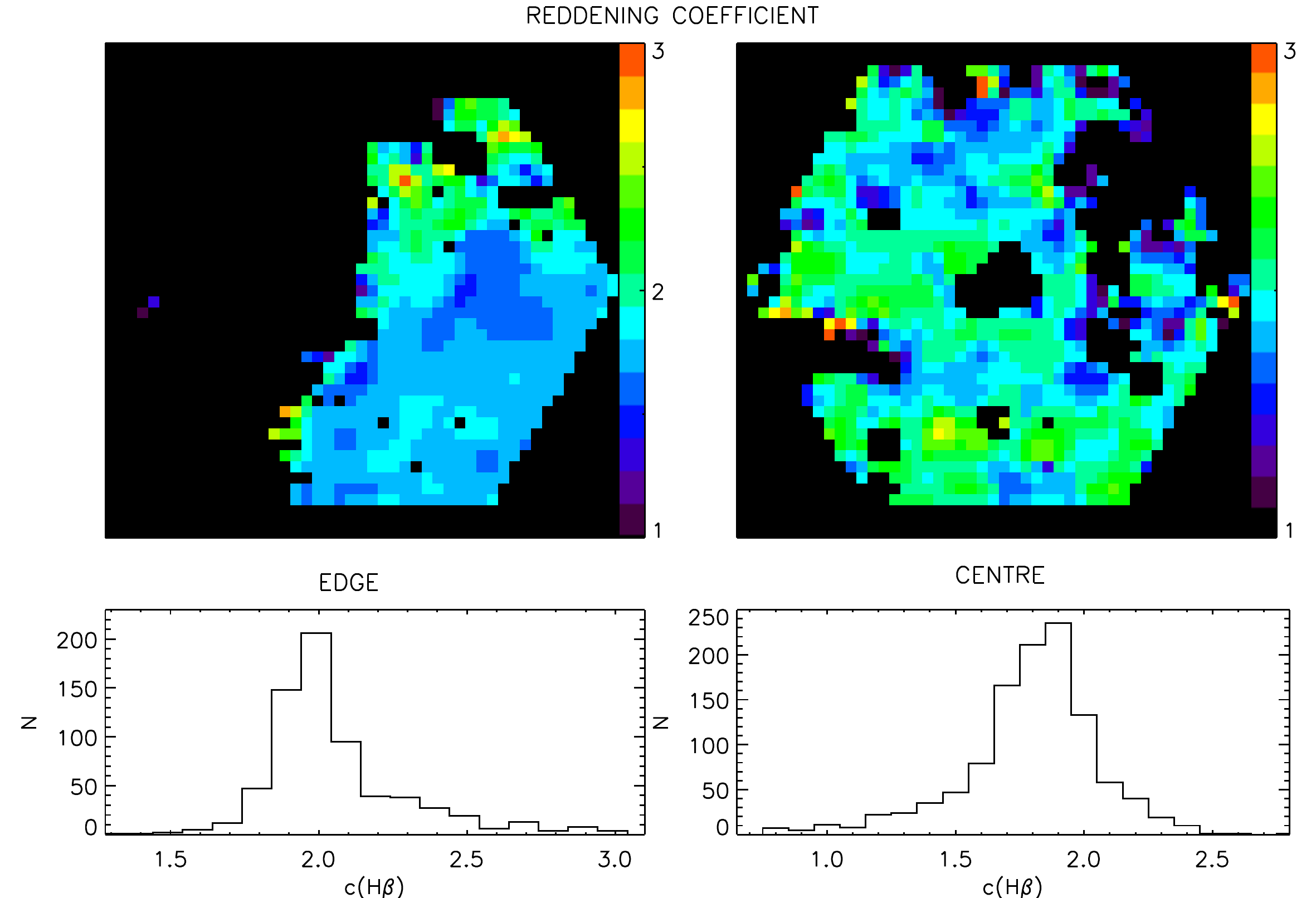}
\caption{Spatial structure of the derived c(H${\beta}$) maps and their corresponding statistical frequency distributions with a binning of 0.1. On the left the edge pointing and on the right the central one. Orientation and sizes of maps are as in Fig. \ref{fig:morphology_all}.}
\label{fig:reddening}
\end{figure}

All the emission line maps were reddening-corrected using the reddening coefficient, c(H${\beta}$), map (each pointing was corrected with its own c(H${\beta}$) map). To determine this coefficient we resorted to the H${\alpha}$/H${\beta}$ line ratio. We analysed the maps of the three Balmer lines detected in our wavelength spectral range (H${\gamma}$, H${\beta}$, and H${\alpha}$) and decided to discard the H${\gamma}$ flux because the S/N was lower than 5 in $\sim$20$\%$ of the spaxels. However, we checked that both derivations were consistent in the spaxels with good S/N. We used an intrinsic Balmer emission line ratio of  H${\alpha}$/H${\beta}$ =3.03 obtained from the public software of \citet{1995MNRAS.272...41S}, assuming Case B recombination with an electron density of $n_{\mathrm{e}}\sim1000$ cm$^{-3}$ \citep{1991A&A...244..205E} and an electron temperature of $T_{\mathrm{e}}\sim 7000$ K (the mean value between the estimations of \citealt{1978ApJ...219..914B} ($\sim$7500 K) and \citealt{1991A&A...244..205E} ($\sim$6000 K)). Statistical frequency distributions of the reddening coefficient were also created for the two maps taking the mean error ($\sim$0.1) as binning. Figure \ref{fig:reddening} shows the spatial distribution of the two derived c(H${\beta}$) maps and their corresponding histograms.

The structure of the reddening map of the central region is mostly uniform with values ranging from 1.3 to 2.5 and a mean value of $\sim$1.85$\pm$0.10; the histogram reveals that the most probable value for c(H${\beta}$) in this zone is 1.90. Single pixels with very high or very low values have a large error in the coefficient estimations. Big holes in the map correspond to the masked pixels.

The derived reddening coefficient map of the edge pointing has a less homogeneous structure, with a c(H${\beta}$) mean value of $\sim$2.11$\pm$0.08 over the 1.7-2.8 range. It is interesting to notice that all the pixels with c(H${\beta}$)$>$2.5 are placed in the NW area, where the discontinuity of the H${\alpha}$ image is observed (Fig. \ref{fig:morphology_all}). In this region, pixels with c(H${\beta}$)$>$2.5 were inspected individually; after checking the S/N of the Balmer lines and the c(H${\beta}$) errors we decided not to mask them and to pay special attention to the rest of properties derived there. We study this region in detail in the 1D analysis (Sect. \ref {1d}). The statistical frequency distribution of the reddening coefficient for the edge pointing gives 2.0 as the most probable value. If we exclude values higher than 2.5, the distribution can be fitted by a Gaussian function. 

To compare our results with the literature, we estimated the extinction as $A_{\mathrm{v}}=2.145 \times c(H{\beta})$, using the \citet{1989ApJ...345..245C} extinction law with $R_{\mathrm{v}}=3.1$ and the colour excess as $E(B-V)=0.692 \times c(H{\beta})$. The mean values derived were $A_{\mathrm{v}}$=3.9 and E(B-V)=1.3 for the central pointing, and $A_{\mathrm{v}}$=4.5 and E(B-V)=1.5 for the edge. The reddening coefficients derived from our data are higher than those estimated by \citet{1991A&A...244..205E} ($\sim$1.35), whereas the $A_{\mathrm{v}}\sim3.8$ obtained by \citet{1981ApJ...249..586C} and $E(B-V)\sim1.35$ from \citet{1975ApL....16..165C} agree with our values.\\

\begin{figure}
\centering
\includegraphics[width=9cm]{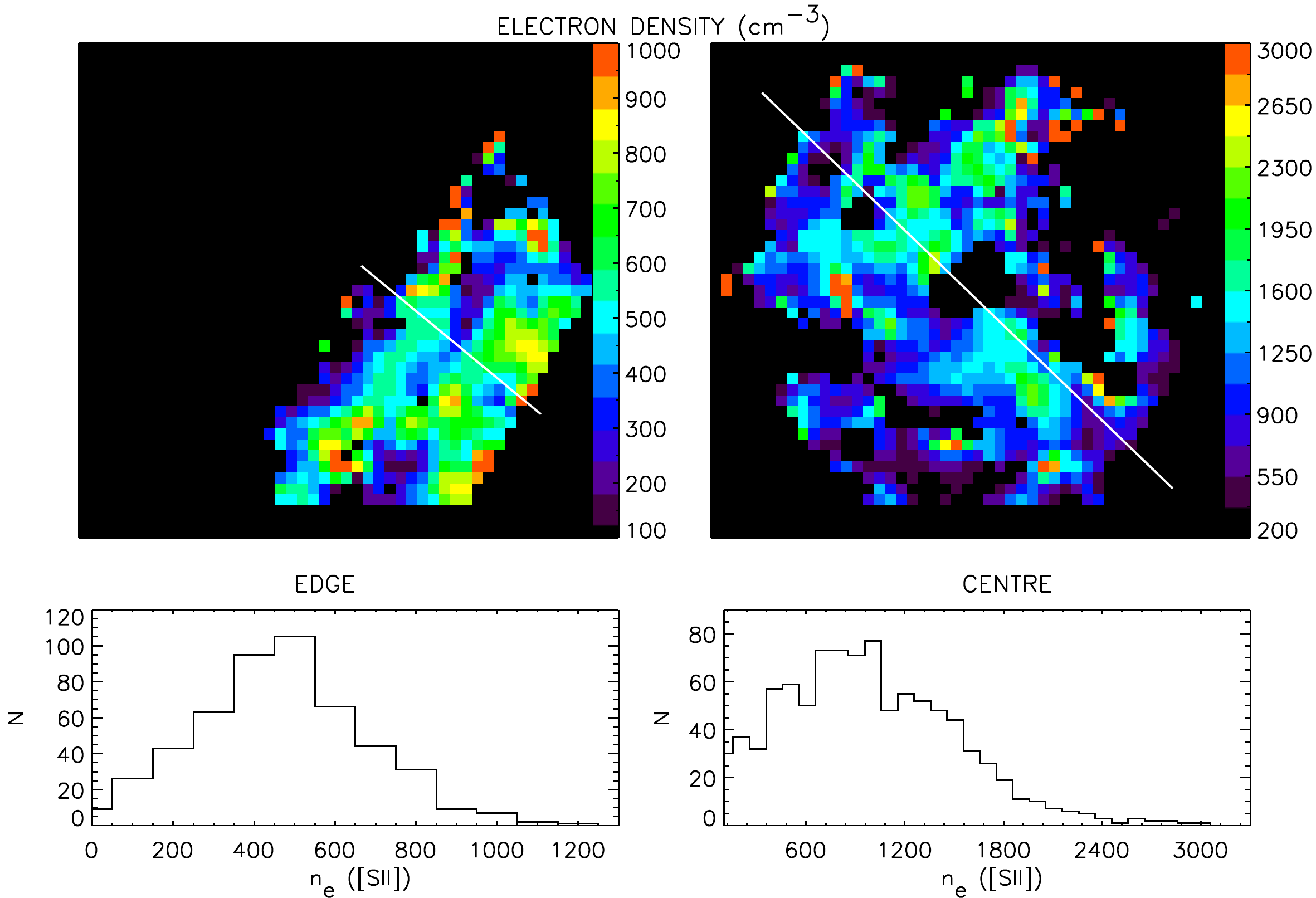}
\caption{Electron density, n$_{\mathrm{e}}$, maps derived from the [S{\sc ii}]$\lambda\lambda$6717/6731 line ratios in units of cm$^{-3}$. Orientations and sizes as in Fig. \ref{fig:morphology_all}. On the bottom the statistical frequency distributions with a binning of 100~cm$^{-3}$. The edge pointing is on the left, and the central on the right. The white lines across the maps (from NE to SW) represent the direction along which the cuts were extracted to study the radial variation of n$_{\mathrm{e}}$. See text for details.}
\label{fig:density}
\end{figure}

The electron density (n$_{\mathrm{e}}$) maps were produced from the [S{\sc ii}]$\lambda\lambda$6717/6731 ratios using the IRAF package TEMDEN based on a five-level statistical equilibrium model \citep{1987JRASC..81..195D,1995PASP..107..896S}. Using these maps, we created statistical frequency distribution of the electron density with a binning of 100~cm$^{-3}$ (low density limit). They are shown with the derived n$_{\mathrm{e}}$ maps in Fig. \ref{fig:density}. In general terms, the values of n$_{\mathrm{e}}$ presented in these maps are in good agreement with values reported in the literature. 

The histogram for the central pointing shows elements distributed in a wide range of density from $\sim$200 to $\sim$3000~cm$^{-3}$ with 1000~cm$^{-3}$ as the most probable n$_{\mathrm{e}}$. The mean value of the distribution is 1008~cm$^{-3}$. Some isolated pixels appear very intense in the image with densities as high as 3000~cm$^{-3}$, but with large errors. The density distribution follows the low-ionization emission elements supporting the idea of a bipolar structure. It is interesting to notice that the knots with the higher surface brightness correspond to the denser zones.

The histogram for the edge pointing shows a distribution close to a Gaussian centred on 500~cm$^{-3}$. The map ranges from 100~cm$^{-3}$ to 1000~cm$^{-3}$ with a mean value of 507~cm$^{-3}$. As happens in the other region, some pixels (7) show higher densities (up to 1000~cm$^{-3}$), and we removed them from our estimations. The majority of the pixels with high reddening coefficient (c(H${\beta}$)$>$2.5) were rejected by the S/N mask for the sulphur line, but the unmasked pixels present a mean density of 613~cm$^{-3}$.

\begin{figure}
\centering
\includegraphics[width=9cm]{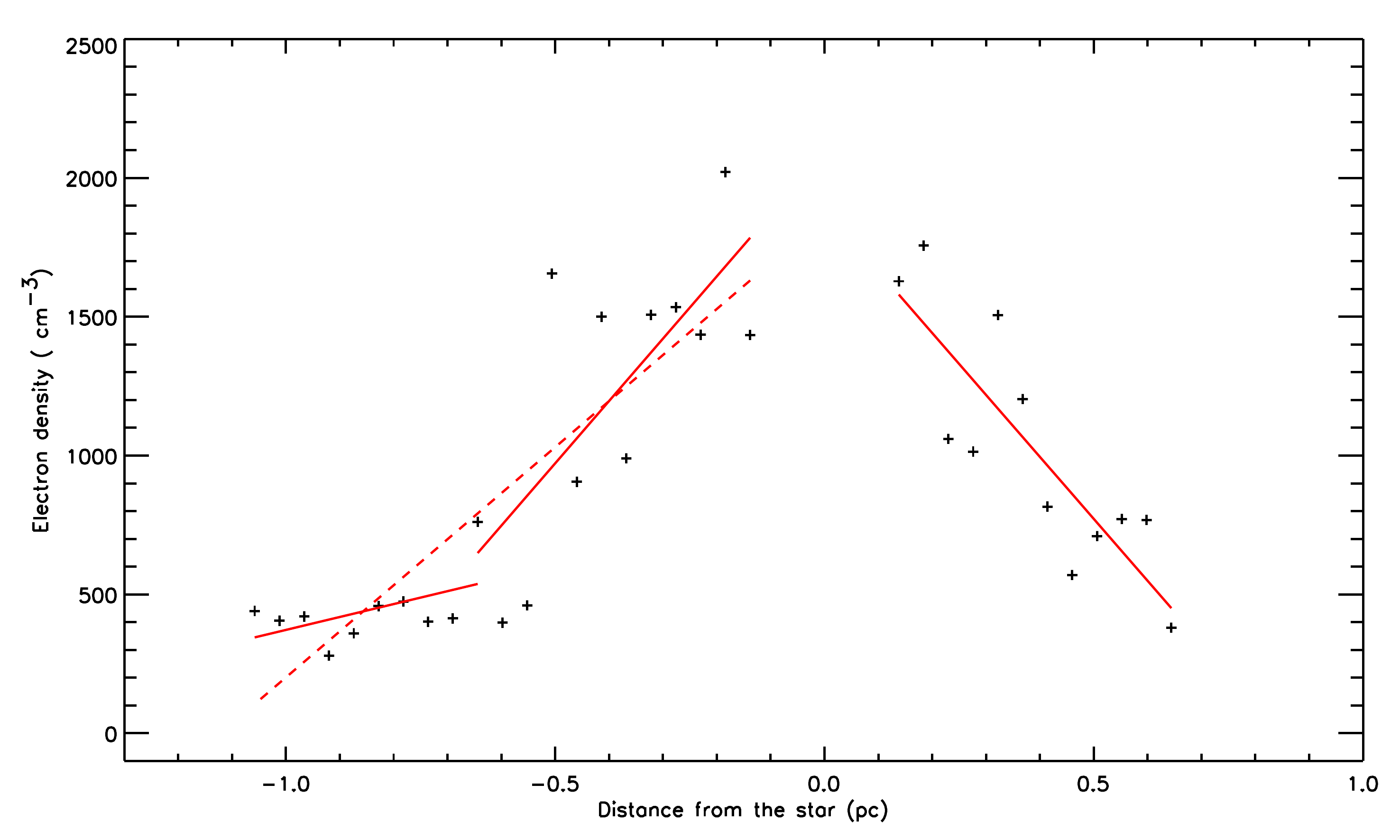}
\caption{Radial variation in electronic density (in cm$^{-3}$) with distance (in pc) along the direction of bipolarity (from NE to SW). We consider negative radius from the star towards the NE and positive from the star towards the SW. Lines indicate least-squares fits: solid lines correspond to pixels differentiating the two pointings, and the dashed line represents the fit along the direction star-NE including pixels from both pointings.}
\label{fig:dens_rad}
\end{figure}

The morphological analyses showed that the bright knots are aligned in a preferred axis along the NE-SW direction with a bipolar structure (see Fig. \ref{fig:morphology_all}); to check that the electron density is related with the bipolarity, we performed a cut in the density maps along this direction (see cuts in Fig. \ref{fig:density}), the density profiles obtained are presented in Fig. \ref{fig:dens_rad}. For them, we performed four fits using the least-squares method: the first from the star towards the SW, the second from the star towards the NE including pixels from the two pointings (centre and edge), and the last two fits from the star towards the NE, but differentiating the two pointings (see Fig. \ref{fig:dens_rad}). It can be seen that the density decreases when we move away from the WR star. In addition, the fits show a symmetric gradient in the central points with a tendency to flatten out towards the ends.

\subsection{Emission line relations \label{diagdiag}}
\begin{figure}
\centering
\includegraphics[width=9cm]{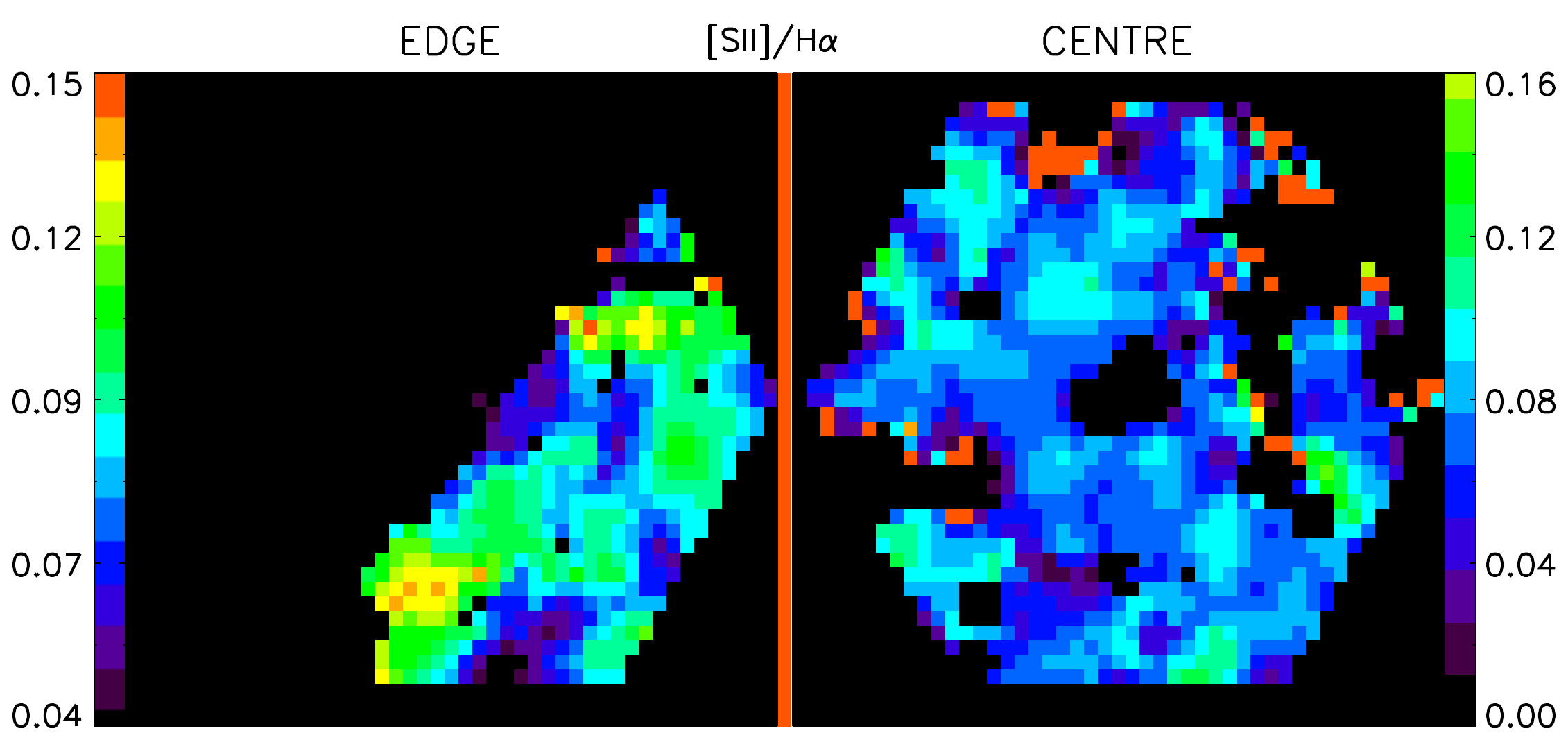}
\includegraphics[width=9cm]{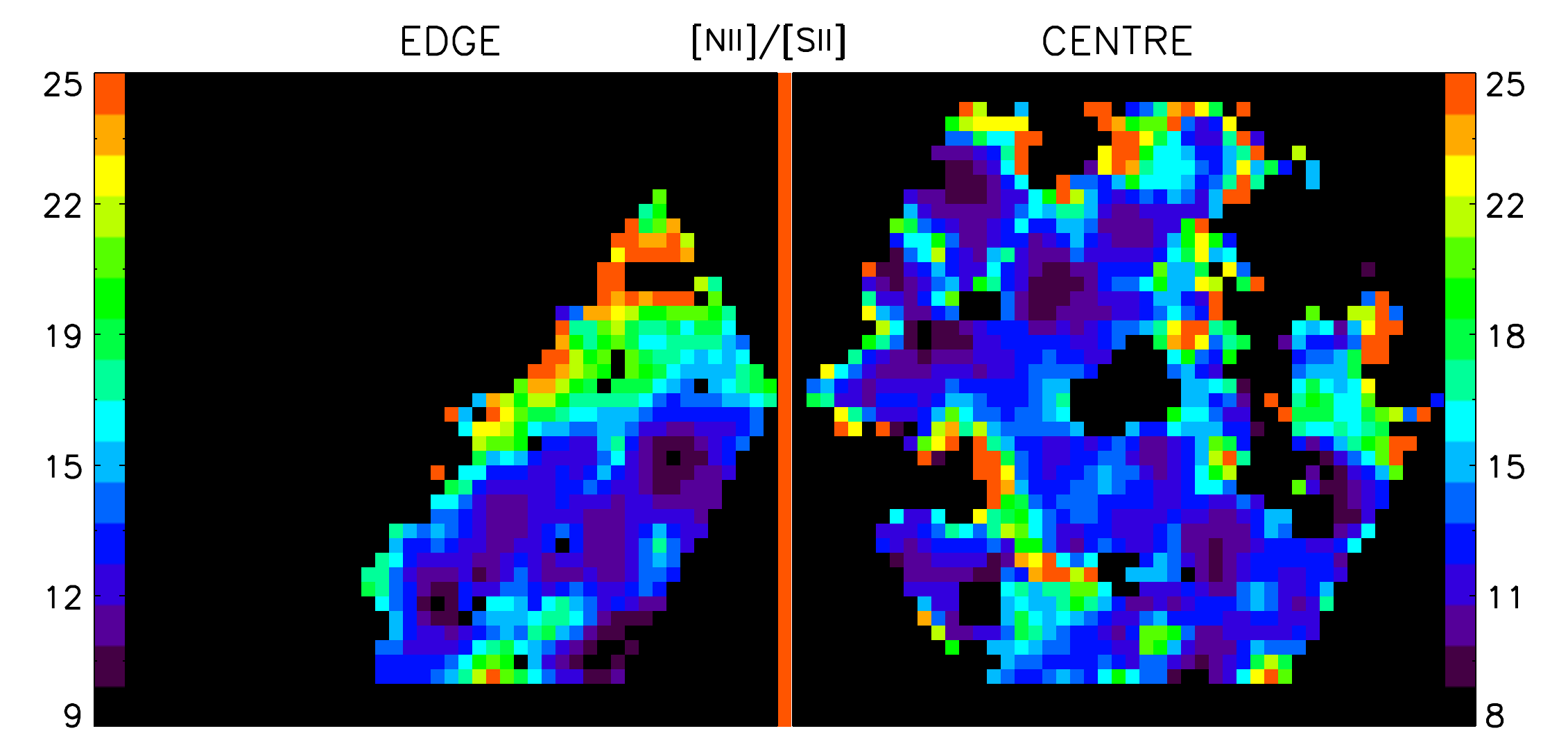}
\includegraphics[width=9cm]{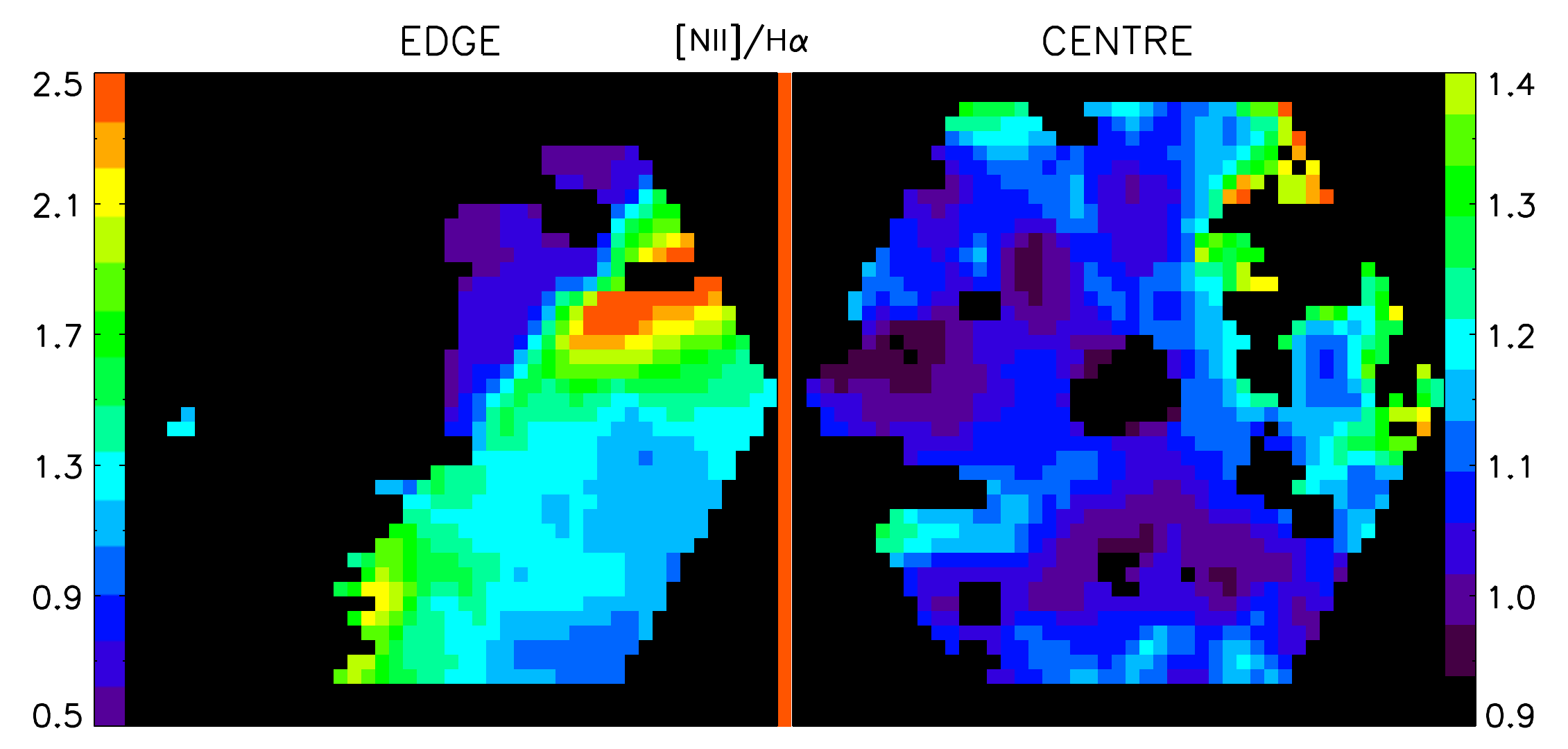}
\caption{Derived maps of the emission line ratios of the two pointings: edge (left) and centre (right). Top: [S{\sc ii}]$\lambda \lambda$6717,6731/H${\alpha}$. Middle: [N{\sc ii}]$\lambda $6584/[S{\sc ii}]$\lambda \lambda$6717,6731. Bottom: ([N{\sc ii}]$\lambda\lambda$6584/H${\alpha}$. Orientations and sizes as in Fig. \ref{fig:morphology_all}.}
\label{fig:ratio_maps}
\end{figure}

Figure \ref{fig:ratio_maps} shows maps of the emission line ratios for the two pointings. Their mean values are summarized in Table \ref{table:line_ratios}. All the intensities presented in the table and figure are reddening-corrected. 

In both regions the [S{\sc ii}]$\lambda\lambda$6717,6731/H${\alpha}$ map presents an inhomogeneus and patchy structure. [S{\sc ii}] lines are fainter than H${\alpha}$ in all the spaxels, with a maximum logarithmic ratio of -1.3 to the north of the edge pointing. Some isolated pixels show higher values, but they are over the limits of the region masked, so unreliable. 

The distribution of the [N{\sc ii}]$\lambda $6584/[S{\sc ii}]$\lambda \lambda$6717,6731 map presents a structure opposite to [S{\sc ii}]/H${\alpha}$ in both regions. The [N{\sc ii}]  emission is stronger than [S{\sc ii}], reaching $\log$([N{\sc ii}]/[S{\sc ii}])=1.4 in areas close to the ISM.

Studying the [N{\sc ii}]$\lambda\lambda$6584/H${\alpha}$ maps led to more interesting results. The central pointing shows positive values, except in some regions in the direction of the bipolarity, where [N{\sc ii}] and H${\alpha}$ fluxes are equal. In the edge pointing, regions with different ratios are clearly separated. In most of the pixels, [N{\sc ii}]$\ge$H${\alpha}$ with an increasing ratio towards the side. To the north, an area can be seen where H${\alpha}$ $>$[N{\sc ii}]; this region possesses the higher derived c(H${\beta})$, and it was masked in the sulphur maps because of its low S/N ($<$5). The NW area has the highest ratio, possibly produced by the contamination of a nearby field star.\\

\begin{table}[!h]
\caption{Mean values of the emission line ratio maps}
\label{table:line_ratios} 
\centering 
\begin{tabular}{l c c}
\hline
Line ratios & Edge & Centre \\
\hline
\hline \\
$\log$([S{\sc ii}]$\lambda\lambda$6717,6731/H${\alpha}$)& -1.03 &  -1.01 \\
$\log$([N{\sc ii}]$\lambda$6584/[S{\sc ii}]$\lambda \lambda$6717,6731)& 1.15 &  1.07 \\
$\log$([N{\sc ii}]$\lambda\lambda$6584/H${\alpha}$)& 0.07 &  0.06 \\
\\
\hline
\end{tabular}
\end{table}

\begin{figure}
\centering
\includegraphics[width=9cm]{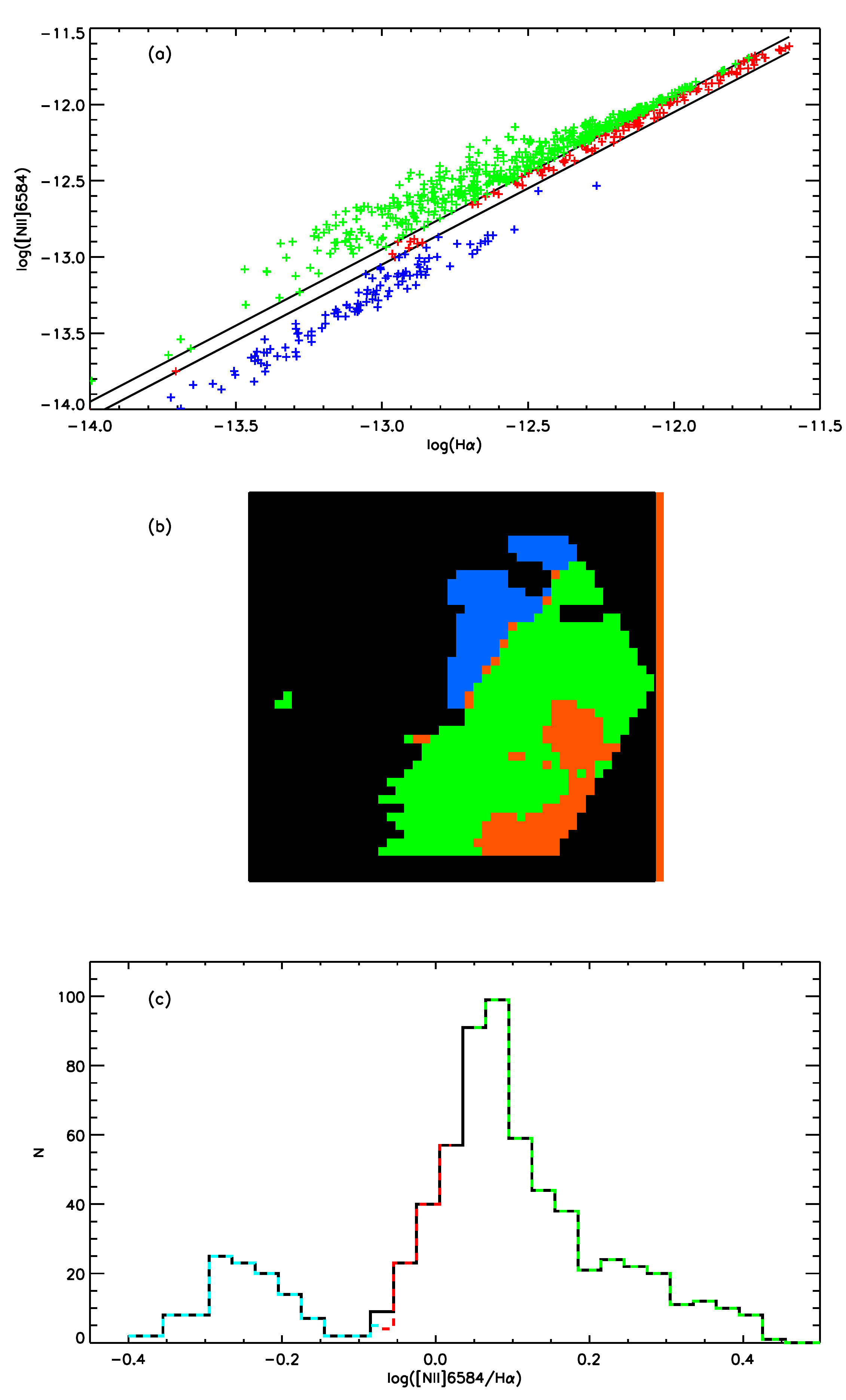}
\caption{Relations between [N{\sc ii}] and H${\alpha}$ for the edge pointing. Colours help us locate points spatially: red corresponds to points with $\log$(H${\alpha}$)=$\log$([N{\sc ii}]) $\pm$ 0.05, blue to points with $\log$(H${\alpha}$)$>\log$([N{\sc ii}]), and green to points where $\log$(H${\alpha}$)$<\log$([N{\sc ii}]). From the top to the bottom: (a) $\log$([N{\sc ii}]$\lambda$6584) vs. $\log$(H${\alpha}$). All the spaxels of the intensity maps (in units of $\log$(erg~cm$^{-2}$~s$^{-1}$) are represented in the diagram with crosses. Black lines with unitary slope represent the limits. (b) PPAK FoV of the edge pointing with the zones defined in plot \emph{a}. (c) Statistical frequency distributions of the $\log$([N{\sc ii}]/H${\alpha}$) map. Black solid line represents the distributions of all the spaxels, and coloured dashed lines represent the regions defined above. See text for details.}
\label{fig:NHa}
\end{figure}

To understand the differences of the [N{\sc ii}]$\lambda\lambda$6584/H${\alpha}$ ratio in the edge pointing, we complement the study by generating statistical frequency distributions of the ratio map and plotting all the spaxels from emission line maps in the diagram [N{\sc ii}]$\lambda$6584 vs. H${\alpha}$ (Fig. \ref{fig:NHa}).The diagram [N{\sc ii}]$\lambda$6584 vs. H${\alpha}$ showed in Fig. \ref{fig:NHa} a, presents double behaviour. We considered two lines with unity slope as upper and lower limits. Points above the upper line are pixels where the [N{\sc ii}] emission is stronger than H${\alpha}$, while below the lower line they have the opposite behaviour. Points between these two lines are pixels with $\log$(H${\alpha}$)=$\log$([N{\sc ii}]) $\pm$ 0.05. Then, we located the points of the diagram in the FoV of PPAK, taking these limits into account, to identify their spatial locations (see Fig. \ref{fig:NHa} b); they appear grouped. The statistical frequency distribution of the [N{\sc ii}]/H${\alpha}$ map shows a bimodal distribution as we can see in Fig.\ref{fig:NHa} c. When we identified the spaxels of the three regions defined above, we found that the left peak (centred in $\sim$-0.3) includes all the points behind the lower limit, and the right peak (centred in $\sim$0.1) includes the points of the two other zones. We can conclude that at least two spatial regions exist in this pointing: one with [N{\sc ii}]$\geq$H${\alpha}$ to the SW and another one to the north with [N{\sc ii}]$<$H${\alpha}$. All the pixels with c(H${\beta}$)$>$2.5 are included in the second region, along with the spectra with very low S/N of the sulphur lines.\\

For the central pointing, the same analysis shows that [N{\sc ii}] follows H${\alpha}$ for all the points in a one-to-one relation line of unity slope. Relations between the other lines were also studied by means of two diagrams ([N{\sc ii}]$\lambda$6584 vs. [S{\sc ii}]$\lambda\lambda$6717,6731 and [S{\sc ii}]$\lambda\lambda$6717,6731 vs. H${\alpha}$), showing strong correlations in both the pointings. The statistical frequency distribution of all the emission line ratios showed single peaks with distributions close to Gaussian functions,  except [N{\sc ii}]$\lambda\lambda$6584/H${\alpha}$ on the edge.

\subsection{The radial velocity field \label{kinematics}}
Limitations on the instrument resolution prevented us from carrying out an exhaustive analysis of the kinematics of M1-67. Nevertheless, the resolution was sufficient for studying the distribution of the radial velocity field and relating it to the morphology and ionization structure.

Using the central wavelength of the Gaussian fit performed in the cubes, we created radial velocity maps for the two observed regions. Two corrections were carried out over the measured radial velocities. First, we estimated the error in the wavelength calibration by comparing the wavelength of a sky emission line with its theoretical value, we obtained a difference of -0.303~\AA{} ($\sim$-16~km~s$^{-1}$ for [OI]$\lambda$5577\AA{}), and this zero point was added to the measured velocities. Then, we translated maps into the local standard of rest (LSR) and corrected for the Earth's motions, taking coordinates and universal time of the observations into account.

\begin{figure}
\centering
\includegraphics[width=8cm]{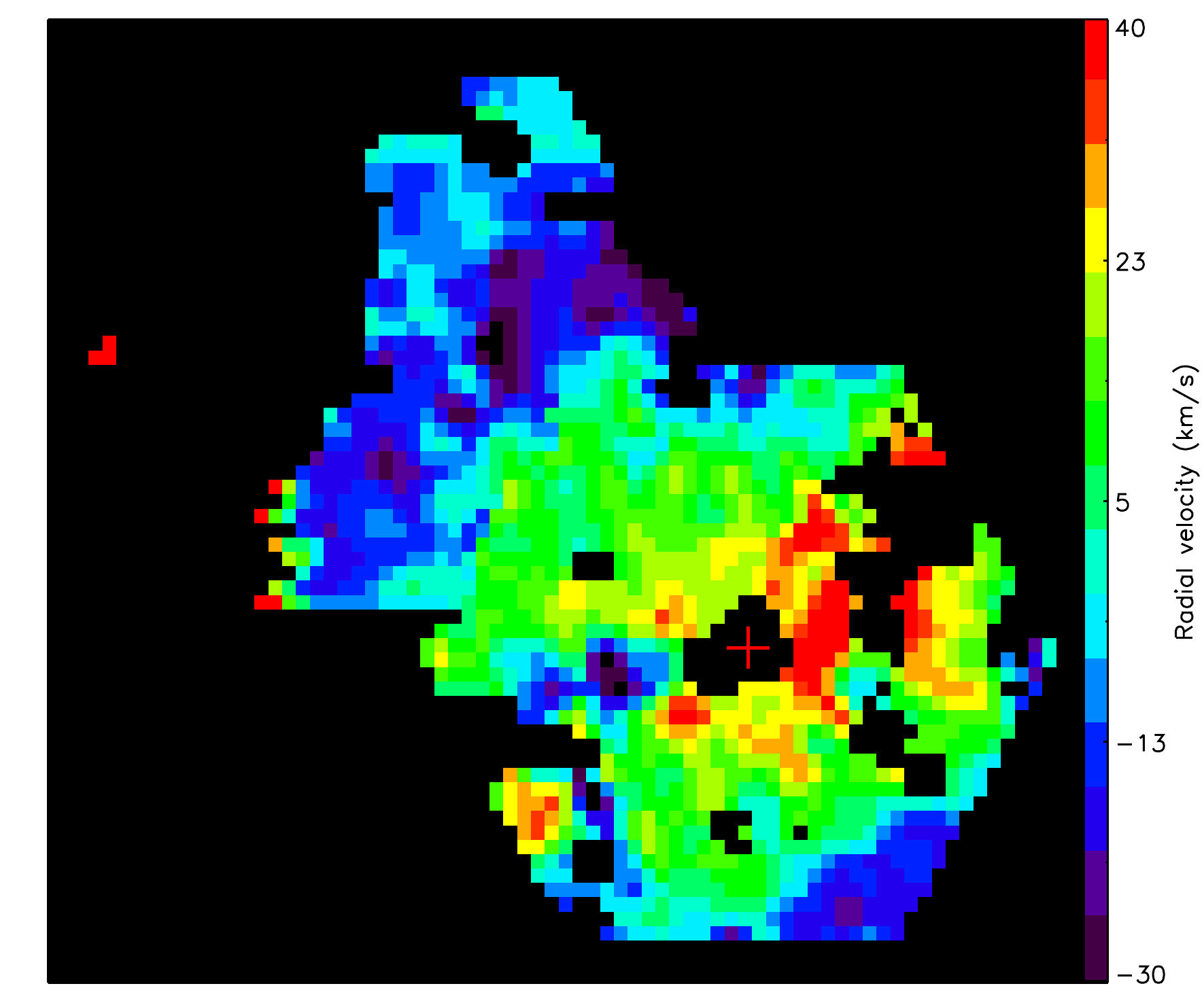}
\caption{Relative radial velocity field derived for H${\alpha}$ in units of km~s$^{-1}$. Zero is the global mean velocity (139~km~s$^{-1}$), see text for details. The two pointings are mosaicked. The red cross marks the position of the central star. North is up and east left (see Fig. \ref{fig:rgb}).}
\label{fig:ha_velocity}
\end{figure}

With the corrected radial velocity fields of H${\alpha}$, we scaled the measured velocities using the overlapping region of the two pointings to avoid deviations. Then, we calculated the total mean velocity, obtaining a value of 139~km~s$^{-1}$, and we established it as the heliocentric velocity of the nebula. This velocity is in very good agreement with the 137~km~s$^{-1}$ obtained by \citet{1998A&A...335.1029S}. We present the relative radial velocity field of H${\alpha}$ for the two regions mosaicked in Fig. \ref{fig:ha_velocity}.\\

Previous kinematic studies \citep{1982A&A...116...54S,1998A&A...335.1029S} have found two components (one redshifted and another blueshifted) supporting the idea of a shell in expansion. With the low resolution of our data we cannot resolve both components, and the velocity field shown is dominated by the radial velocity of the brightest knots, a kind of intensity- weighted radial velocity distribution. Despite the low resolution, a study of the overall structure of both regions can be carried out. The gas of the nebula seems to move faster near to the WR star, decreasing its relative velocity when moving away from the centre. The velocity field changes its tendency (increasing) in the \textquotedblleft peculiar\textquotedblright ~zone towards the north of the edge pointing, where other properties were also found to differ from the rest of the nebula.

\begin{figure}
\centering
\includegraphics[width=9cm]{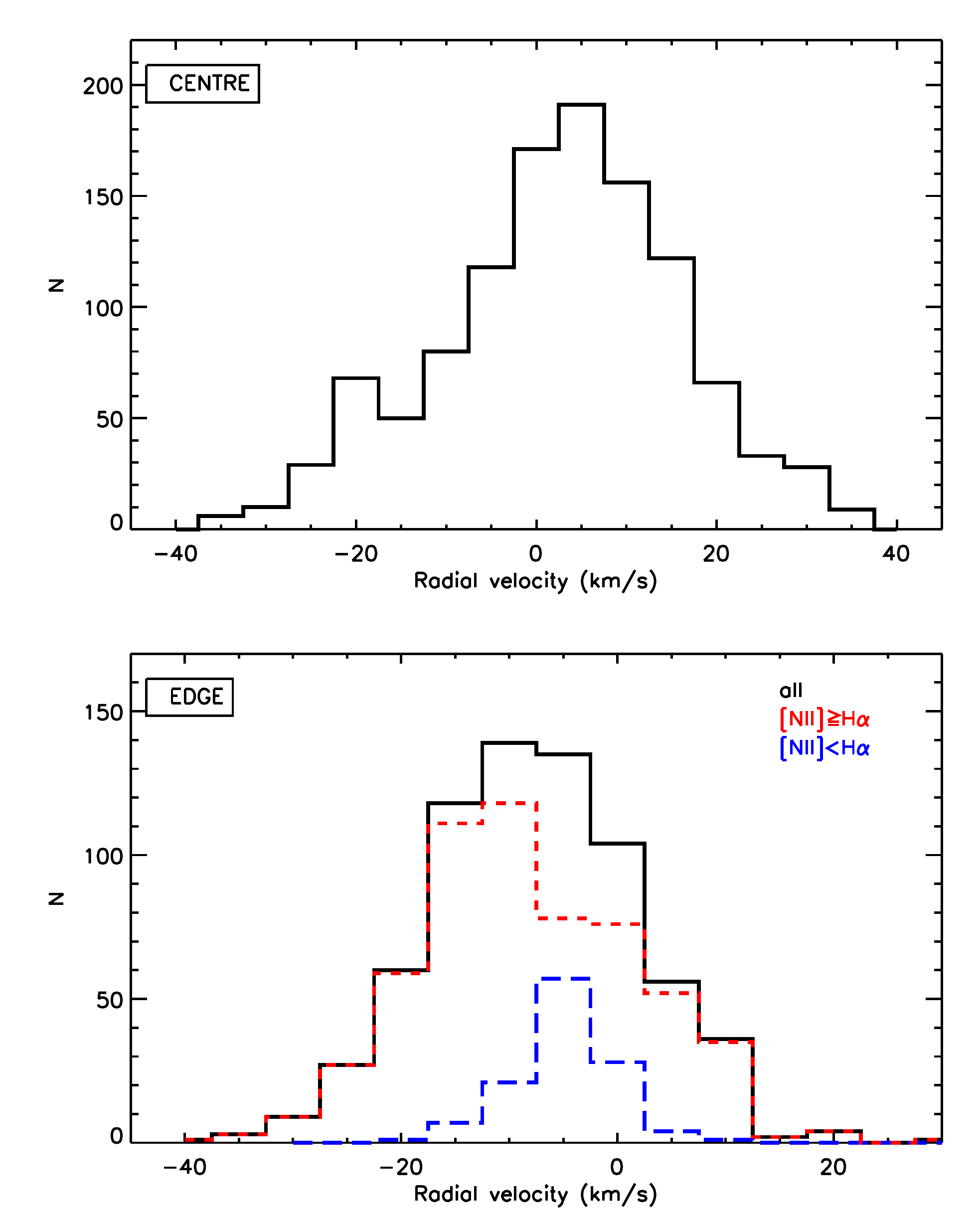}
\caption{Statistical frequency distributions of the radial velocity of H${\alpha}$ relative to the heliocentric velocity, with a binning of 5~km~s$^{-1}$. The central region is on the top with all the spaxeles represented. On the bottom the edge pointing: black solid line represents all the pixels, the short-dashed red line is the region where [N{\sc ii}]$\geq$H${\alpha}$, and the long-dashed blue line represents the regions with [N{\sc ii}]$<$H${\alpha}$.}
\label{fig:hist_vr}
\end{figure}

Figure \ref{fig:hist_vr} shows the statistical frequency distributions of the radial velocity maps with a binning of 5~km~s$^{-1}$. To consolidate the differences found in the diagram [N{\sc ii}]$\lambda$6584 vs. H${\alpha}$ for the edge pointing (Fig. \ref{fig:NHa}), we represented pixels from the two regions separately  The region where [N{\sc ii}]$\geq$H${\alpha}$ presents a Gaussian distribution that covers a wide range in velocity, suggesting that some regions are moving away from us and others towards us. The distribution of the regions where H${\alpha}$ emission is higher than [N{\sc ii}] is narrower and it is centred near to the zero velocity.

\section{Properties of the integrated spectra \label{1d}} 

\begin{figure}
\centering
\includegraphics[width=9cm]{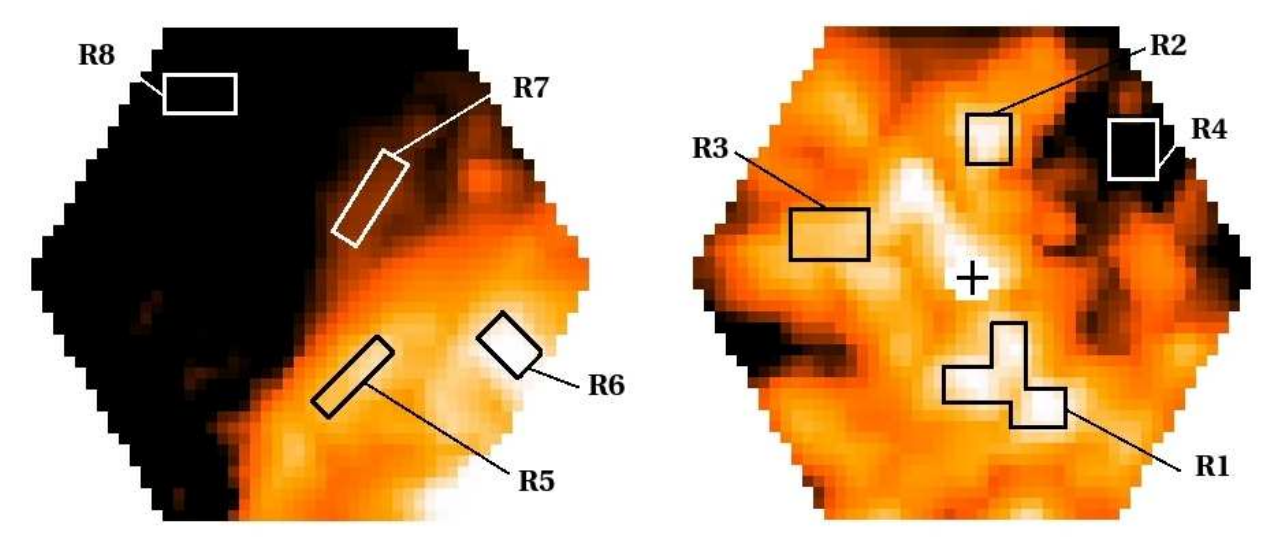}
\caption{H${\alpha}$ images of the two areas of M1-67 observed with PPAK. Boxes represent the eight regions where the integrated spectra were generated. For the offsets of each region from the central star (green cross), see the text. Orientations and sizes are as in Fig. \ref{fig:morphology_all}. Edge on the left and centre on the right.}
\label{fig:integratedregions}
\end{figure}

We created 1D spectra by combining fibres to describe the integrated properties of several interesting zones. Eight integrated spectra were generated over the two pointings (see Fig. \ref{fig:integratedregions}).\\

The regions selected over the central pointing are:

-\textit{Region 1} (R1): examining the emission of the low ionization elements showed in Fig. \ref{fig:morphology_all}, three bright knots appear to the south of the nebula. Eight fibres over these knots were selected and combined to create a single spectrum. The offset from central star is $\Delta\alpha\sim$4.05$\arcsec$, $\Delta\delta\sim$13.5$\arcsec$.

-\textit{Region 2} (R2): we combined three spaxels to the north of the star coinciding with another isolated knot. The offset from central star is $\Delta\alpha\sim$1.35$\arcsec$, $\Delta\delta\sim$14.85$\arcsec$.

-\textit{Region 3} (R3): we chose those fibres placed to the east of the star in a zone where an extended emission is seen in H${\alpha}$, paying attention to not include light from any star. The offset from central star is $\Delta\alpha\sim$12.15$\arcsec$, $\Delta\delta\sim$4.05$\arcsec$.

-\textit{Region 4} (R4): we were interested in analysing a large region in the NW masked in the 2D analysis where all the emission line maps showed S/N lower than 5. The fourth integrated spectrum was created there to check that there is emission in this area. The offset from central star is $\Delta\alpha\sim$14.85$\arcsec$, $\Delta\delta\sim$14.85$\arcsec$.\\

The regions selected over the edge pointing are:

-\textit{Region 5} (R5): nine fibres were selected on the south of the edge pointing, close to the discontinuity. This spectrum belongs to the region showed in Fig. \ref{fig:NHa} b where [N{\sc ii}] is stronger than H${\alpha}$. The offset from central star is $\Delta\alpha\sim$31.05$\arcsec$, $\Delta\delta\sim$16.2$\arcsec$.

-\textit{Region 6} (R6): we combined several spaxels at the SW limit of the FoV to check that [N{\sc ii}]$\sim$H${\alpha}$ in this area, as we found in Sect. \ref{diagdiag}. The offset from central star is $\Delta\alpha\sim$14.85$\arcsec$, $\Delta\delta\sim$20.25$\arcsec$.

-\textit{Region 7} (R7): in a faint region to the north of the edge pointing where interesting properties were obtained in the 2D analysis: some pixels show c(H${\beta}$)$>$2.5, the S/N of the sulphur lines is very low so they were masked in several maps, the [N{\sc ii}]/H${\alpha}$ ratio has its minimum values, and the kinematic study revealed that here the radial velocity increases opposite to the general trend. Seven spaxels were combined in this region to analyse the properties in detail. The offset from central star is $\Delta\alpha\sim$27$\arcsec$, $\Delta\delta\sim$40.5$\arcsec$.

-\textit{Region 8} (R8): six fibres were selected on the left of the discontinuity to checked whether this region has nebular emission. The offset from central star is $\Delta\alpha\sim$47.25$\arcsec$, $\Delta\delta\sim$52.65$\arcsec$.\\

\begin{figure*}
\centering
\includegraphics[width=\textwidth]{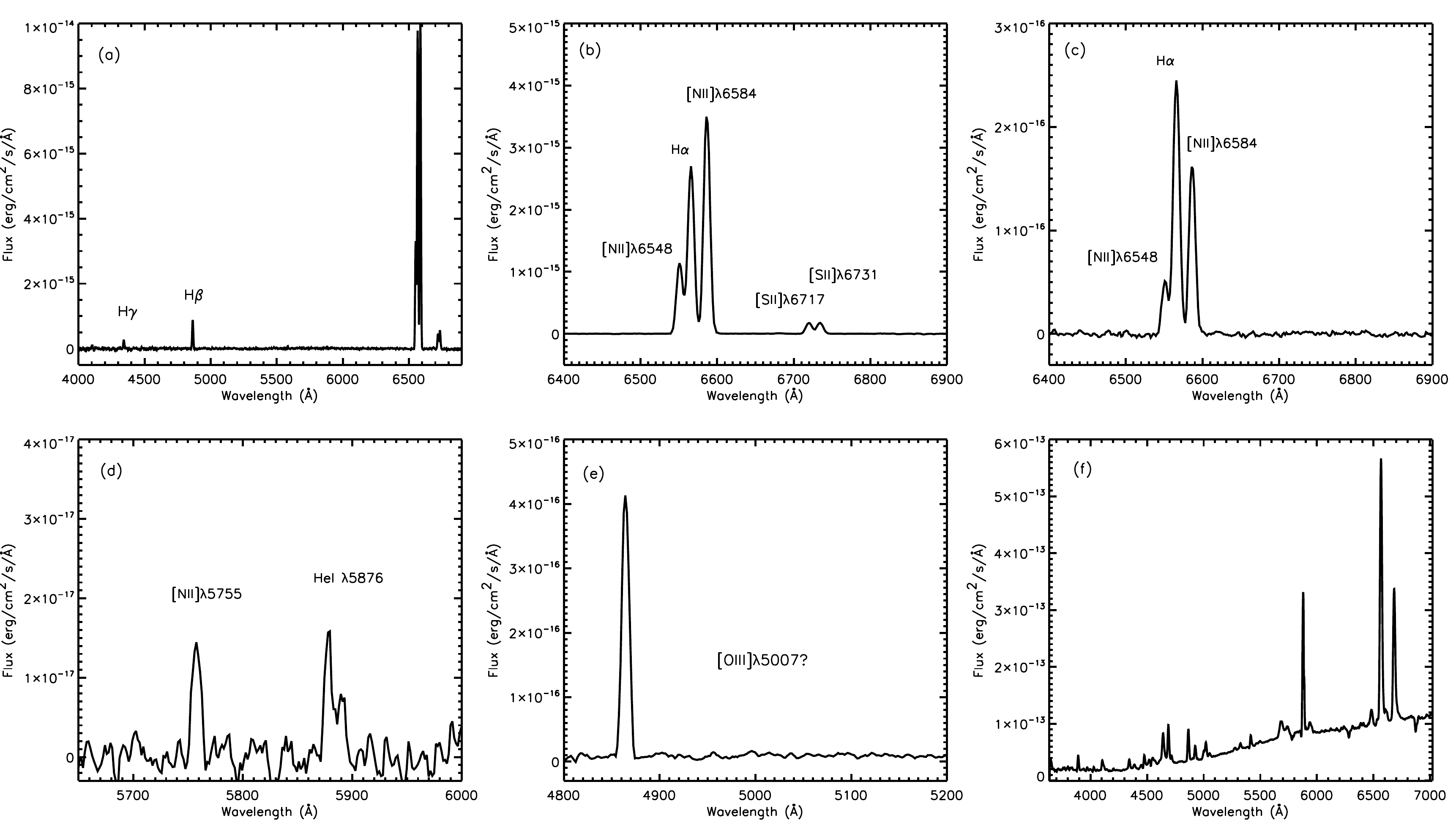}
\caption{Examples of integrated spectra. From left to right and top to bottom: (a) Whole spectrum of \textit{Region 3}. (b) Spectrum of \textit{Region 5} in the range of H${\alpha}$. (c) \textit{Region 7}, in same range as \textit{b}, where the absence of sulphur lines can be seen. (d) Spectrum of \textit{Region 5} centred on the [N{\sc ii}]$\lambda$5755\AA{} emission line, used to calculate the electron temperature. (e) Zoom over the spectrum in \textit{Region S2} without any emission in the [O{\sc iii}]$\lambda$5007\AA{} emission line.(f) Whole spectrum of WR124 obtained from the central spaxel. }
\label{fig:integratedspectra}
\end{figure*}

In addition, another three integrated spectra were extracted to perform several tests. From the central spaxel of the central pointing FoV, we obtained the spectrum of WR124 (\textit{Region WR}). The other two were extracted at $\sim$15\arcsec\ to the NE of the star (common region in both pointings) to study the zone where \citet{1981ApJ...249..586C} found emission in [O{\sc iii}]$\lambda$5007\AA{} (\textit{Regions S1 and S2} in the central and edge pointings, respectively). Figure \ref{fig:integratedspectra} shows six representative 1D spectra from the 11 created. \\

Fluxes of the main emission lines were measured by fitting Gaussian functions using SPLOT within IRAF. All the measured fluxes are in units of erg~cm$^{-2}$~s$^{-1}$ per fibre (area of fibre $\sim$5.7~arcsec$^2$). Statistical errors were estimated using the formula presented  in \citet{2003MNRAS.346..105P}:
\begin{equation}
\sigma_{\mathrm{1}}=\sigma_{\mathrm{c}} N^{1/2} [1+EW/(N\Delta)]^{1/2}
\end{equation}
where $\sigma_{\mathrm{1}}$ represents the error on the observed line flux, N is the number of pixels used to measure the line, EW the line equivalent width, $\sigma_{\mathrm{c}} $ the standard deviation in the continuum close to the line of interest, and $\Delta$ represents the dispersion in \AA{}/pix.\\

We derived the reddening coefficient, c(H${\beta}$), from the H${\alpha}$/H${\beta}$ and H${\gamma}$/H${\beta}$ line ratios using the procedure described in Sect. \ref{maps}. In \textit{Region 7}, H${\gamma}$  was measured with low S/N, and only the other two Balmer lines were used to estimate c(H${\beta}$). The reddening coefficients agree with values obtained in the 2D study. Table \ref{table:all_lines} lists the reddening-corrected fluxes of the emission lines measured in every zone labelled with their standard identification. The third column reports the adopted reddening curve using extinction law by \citet{1989ApJ...345..245C} with $R_{\mathrm{V}}=3.1$.  Errors in the emission line intensities were derived by propagating the observational errors in the fluxes and the reddening constant uncertainties. The estimated fluxes and errors were normalized to $F(H{\beta})=100$. The values obtained for c(H${\beta}$) are also presented in the last row of Table \ref{table:all_lines}. \\

Five integrated spectra deserve special attention. First, the R4 created in the NW dark area of the central pointing only showed three emission lines: H${\alpha}$, [N{\sc ii}]$\lambda$6548\AA{}, and [N{\sc ii}]$\lambda$6584\AA{}. We deduce that, in areas out of the bipolar structure a faint, but not negligible, emission exists coming from the nebular gas rather than the ISM. We estimated the H${\beta}$ flux by means of the reddening coefficient: assuming c(H${\beta}$)=1.87 (the mean value of the other integrated spectra of this pointing), we performed the inverse process of the extinction correction and obtained F($H{\beta}$)=3.21~10$^{-16}$~erg~cm$^{-2}$~s$^{-1}$. 

The spectrum of R8 does not show any emission, thus physical and chemical properties could not be estimated here. We did not include this region in tables. We extracted a spectrum of the WR star (\textit{Region WR}) shown in Fig.\ref{fig:integratedspectra}f, and not perform a detailed analysis here. Finally, the study performed over \textit{Regions S1 and S2} revealed typical nebular spectra (very similar to R4), but we did not find any emission of the [O{\sc iii}]$\lambda$5007\AA{} line as seen in Fig. \ref{fig:integratedspectra}e.

\subsection{Physical properties and chemical abundances \label{prop_and_ab}}
Electron density (n$_{\mathrm{e}}$) was calculated from the [S{\sc ii}]$\lambda\lambda$6717/6731 line ratio using the IRAF package TEMDEN. The derived density ranges from $\sim$1500~cm$^{-3}$ near the star, to $\sim$650~cm$^{-3}$ towards the edge. These values are consistent with our 2D maps and with previous studies \citep{1991A&A...244..205E,1998A&A...335.1029S}. \\

Electron temperature, T$_{\mathrm{e}}$, can be derived using the line ratio R$_{\mathrm{N2}}$: 
\begin{equation}
R_{\mathrm{N2}}={I([\mathrm{N \textsc{ii}}]\lambda 6548)+I([\mathrm{N \textsc{ii}}]\lambda 6584) \over I([\mathrm{N \textsc{ii}}]\lambda 5755)}.
\end{equation}
The [N{\sc ii}]$\lambda$5755\AA{} auroral line that appears close to the \textquotedblleft sky\textquotedblright\, line Hg{\sc i} 5770\AA{}, was detected in two zones (R5 and R6). We measured this line again in the spectra before sky subtraction and conclude that the flux of [N{\sc ii}]$\lambda$5755\AA{} line in R6 is contaminated by the Hg{\sc i} emission line, thus not reliable. We obtained a direct estimate of T$_{\mathrm{e}}$([N{\sc ii}]) from R$_{\mathrm{N2}}$ only for R5.\\

To reinforce the validity of the chemical abundances estimations and to provide ionization correction factors (ICFs) for those species whose ionizations stages were not all observed in the optical spectrum, we performed photoionization models of R5. To do so, we used the code CLOUDY v.10 \citep{1998PASP..110..761F}, assuming a central ionizing source from a WR star atmosphere \citep{2002MNRAS.337.1309S} with $Z = 0.008$ and an effective temperature of the star of 45~000 K which are, respectively, the closest values to the measured total metallicity of the gas and the estimated temperature of WR124 \citep{2006A&A...457.1015H}. 

We considered a spherical geometry putting the gas at a distance of 1~pc from the star and assumed a constant density of 700~cm$^{-3}$, a value similar to the one derived from [S{\sc ii}] emission lines. The model that fits the emission line intensities of [O{\sc ii}], [O{\sc iii}], He{\sc i}, [N{\sc ii}], and [S{\sc ii}]  better in R5 was obtained by varying the ionization parameter (U) and the relative chemical abundances of He, O, N, and S. The emission lines from this model are listed in Table \ref{table:all_lines}, while the derived physical properties and the ionic and total chemical abundances are listed in Table \ref{table:paramyabun}. The ICFs obtained were ICF(N$^{+}$)=1.21 and ICF(S$^{+}$)=1.58. Regarding the resulting geometry, the final radius is 1.22~pc, which is of the same order of magnitude as the apparent size of the nebula in the images. \\

To estimate chemical abundances, electron density and electron temperature are required. We used T$_{\mathrm{e}}$([N{\sc ii}]) as temperature representative of the low ionization ions, S$^{+}$, N$^{+}$, and O$^{+}$, and T$_{e}$([O{\sc iii}]) for  deriving the O$^{2+}$ and He$^{+}$ abundances. In those zones where the electron temperature was not calculated, we adopted the value of R5. In previous studies, T$_{\mathrm{e}}$([N{\sc ii}]) ranges from 5900~K \citep{1998A&A...335.1029S} to 8000~K \citep{1981ApJ...249..586C}; maybe, the supposition of T$_{\mathrm{e}}$=8200~K leads to our abundances being underestimated.  Since the photionization model predicts T$_{\mathrm{e}}$([N{\sc ii}])$\sim$8550~K and T$_{e}$([O{\sc iii}])$\sim$8330~K in R5, we considered T$_{\mathrm{e}}$([N{\sc ii}])$\sim$T$_{e}$([O{\sc iii}]) in the estimations. To infer abundances in R4 and R7, where sulphur lines were not measured, we adopted the electron density of R5, n$_{e}$=631~cm$^{-3}$. We checked that variations in density do not affect this estimation. \\

Ionic abundances were derived from the forbidden-to-hydrogen emission line ratios using the functional forms given by \citet{2008MNRAS.383..209H}, which are based on the IRAF package IONIC. We used equations from \citet{2004ApJ...617...29O} to obtain the singly ionized helium abundance. To determine the total abundance of O/H we added the two ionic abundances (O/H~$\sim$ O$^{+}$/H$^{+}$ + O$^{2+}$/H$^{+}$). The total N/H and S/H abundances were inferred thanks to the ICFs obtained in the photoionization model, X/H$\sim$(X$^{+}$/H) $\times$ ICF(X$^{+}$). In the case of helium abundances we used the relation between X(O$^{2+}$)=O$^{2+}$/(O$^{2+}$+O$^{+}$) and ICF(He$^{+}$+He$^{++}$) from \citet[Fig. 7]{2007ApJ...662...15I} and we deduced that ICF(y$^{+}$)$\gg$1. Since our helium measurements are uncertain, we do not venture to estimate the total helium abundances.\\

In R5 all the useful emission lines were measured and the abundances determined as explained above. In the rest of the regions, we did not measure all the necessary lines to calculate abundances, and we resorted to the empirical parameter N2S2 \citep{2009MNRAS.398..949P} to estimate N/O from the nitrogen and sulphur emission lines:

\begin{equation}
\log(N/O)=1.26\times N2S2 - 0.86
\end{equation}
where

\begin{equation}
N2S2=\log \left ({I([\mathrm{N \textsc{ii}}]\lambda 6584) \over I([\mathrm{S\textsc{ii}}] \lambda\lambda 6717,6731)} \right).
\end{equation}

Before, we estimated the N/O in R5 with the N2S2 parameter and checked that the result was in good agreement with the value obtained with the direct method. In Table \ref{table:paramyabun} we present all the ionic and total abundances, with their corresponding errors, derived for the integrated spectra. We discuss the results in Sect. \ref{chemical}.

\section{Infrared study \label{ir}}
To enhance the morphological and chemical analysis, a study in the mid-infrared was performed. We obtained IRS (Infrared Spectrograph, \citealt{2004ApJS..154...18H}) data in mapping mode and the MIPS (Multiband Imaging Photometer, \citealt{2004ApJS..154...25R}) 24$\,\mu$m image from the Spitzer Heritage Archive (SHA)\footnote{Website: sha.ipac.caltech.edu/applications/Spitzer/SHA}. M1-67 has already been studied in the past in the infrared range by \citet{1985A&A...145L..13V}. They presented the energy distribution of the central star WR\,124 and flux densities, finding thermal emission of dust at T$_{\mathrm{c}}\sim$100~K.\\

\begin{figure}
\centering
\includegraphics[width=9cm]{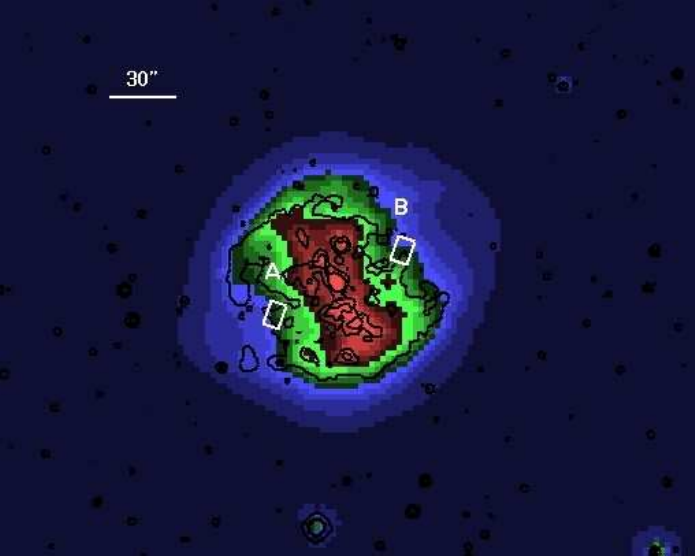}
\caption{MIPS 24$\mu$m image of M1-67. North is up and east left. Boxes indicate the two regions where IR spectra were obtained. Contourns represent the H${\alpha}$ emission derived from Fig. \ref{fig:rgb}.}
\label{fig:spitz_24micr}
\end{figure}

Figure \ref{fig:spitz_24micr} shows the MIPS 24$\mu$m image of M1-67. This image has already been presented by \cite{2010MNRAS.405.1047G}. In a nebula, the origin of the 24$\mu$m emission can be mainly due to two factors: presence of the [O{\sc iv}]25.90$\mu$m line from highly ionized gas or warm dust. Since M1-67 presents a low degree of ionization, we deduce that the observed emission shown in Fig. \ref{fig:spitz_24micr} traces the warm dust distribution of the nebula. The emission has an elliptical shape along the NE-SW direction in very good agreement with the bipolar axis observed in Fig. \ref{fig:morphology_all}, thus suggesting that the structure is composed of a mixture of ionized gas and warm dust. Furthermore, an external and spherical structure can be seen extending around the ellipsoidal shell. This faint bubble is not seen in our optical images.\\

\begin{figure}
\centering
\includegraphics[width=9cm]{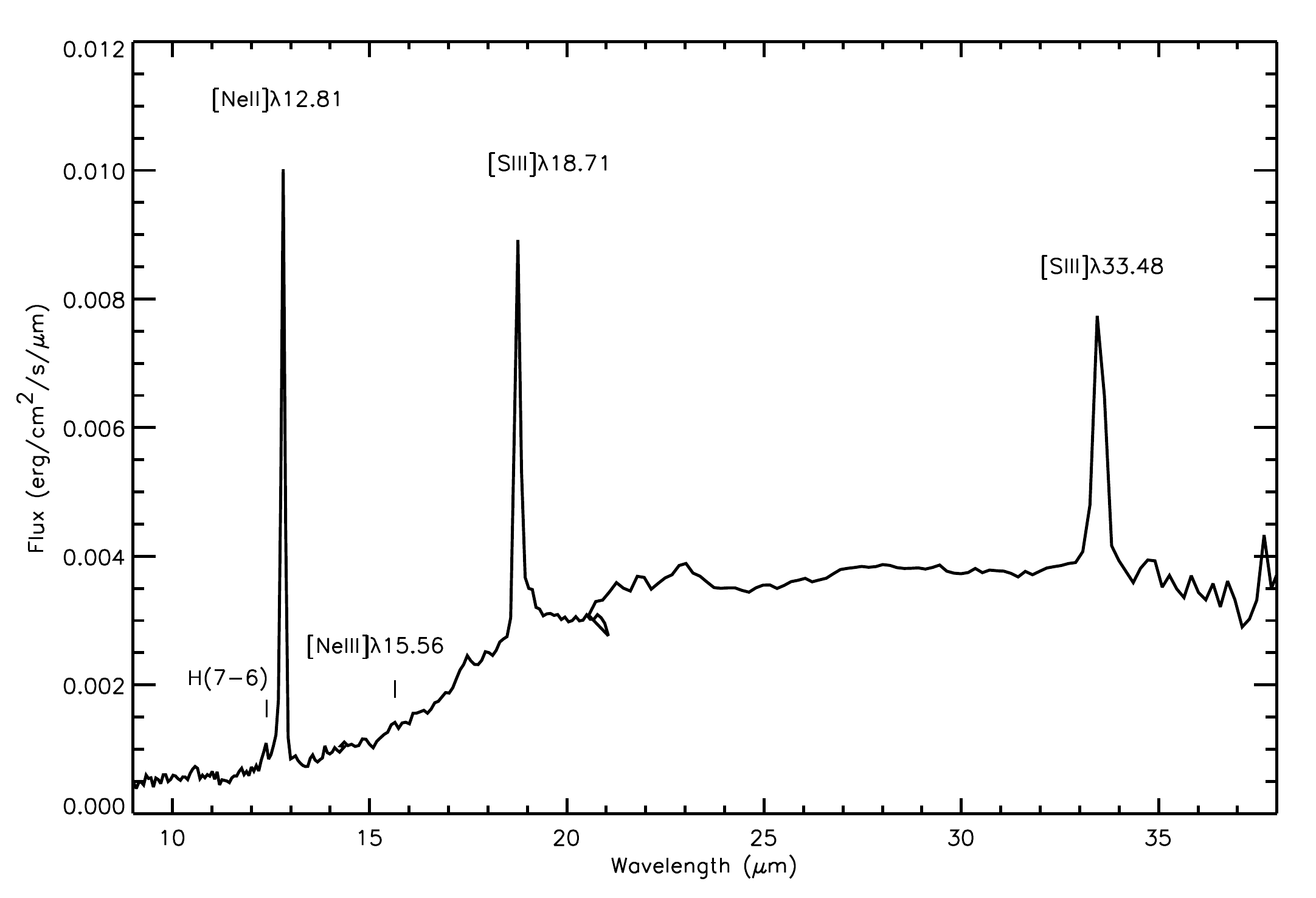}
\caption{Infrared spectrum obtained in Region B. The most relevant lines are indicated.}
\label{fig:spitz_spec}
\end{figure}

For the low-resolution short-low (SL) and long-low (LL) modules spectroscopic observations, basic calibrated data (BCD, pipeline version 18.18) were processed and analysed with the CUBISM software \citep{2007ApJ...656..770S}. Data were background-subtracted using averaged off-source observations and flux-calibrated. Bad pixels were removed with the automatic correction routine within CUBISM and a datacube assembled for each module. CUBISM allows extracting of 1D spectra over polygonal apertures: given the different spatial coverage of the SL and LL module, we chose two apertures (with an area of $\sim$60~arcsec$^2$) on the outskirts of the nebula, observed by both modules. The spectra from the different modules were stitched together, ignoring the noisy region at the red end of each order. We called them Regions A and B (see Fig. \ref{fig:spitz_24micr}). In Fig. \ref{fig:spitz_spec} we present the spectrum obtained in Region B.\\

We measured the most important lines by fitting Gaussian functions with IRAF. Errors were calculated as we explained before (Sect. \ref{1d}). Fluxes and their corresponding errors are presented for the two regions in Table \ref{table:infrared}.\\

Assuming the electron temperature of \textit{Region 5} (T$_{e}$=8200~K) and an electron density n$_{e}$=600~cm$^{-3}$, we inferred the chemical abundances. To obtain the fluxes relatives to H${\beta}$ we used the theoretical ratio of H(7-6)/H${\beta}$=0.0109 from \citet{1995MNRAS.272...41S}. The ionic abundances, Ne$^{+}$/H$^{+}$, Ne$^{2+}$/H$^{+}$, S$^{2+}$/H$^{+}$, and S$^{3+}$/H$^{+}$ were inferred by using the IRAF package IONIC. We estimated the total neon abundance by adding the two ionic abundances, Ne/H$\sim$Ne$^{+}$/H$^{+}$+Ne$^{2+}$/H$^{+}$. For deriving the total S/H abundance we need to add the S$^{+}$/H$^{+}$ from the optical spectra. To do so we compared the regions from which the IR and the optical 1D spectra were taken. Noticing the proximity between Region A and R3, we approximated the total sulphur abundance in the spectrum A as S/H$\sim$(S$^{+}$/H$^{+}$)$_{R3}$ + (S$^{++}$/H$^{+}$)$_{A}$. The 1D spectrum nearest to B is R4, but in R4 we did not measure sulphur lines. Since the S$^{+}$/H$^{+}$ is similar in all the integrated spectra, we considered the mean value, (S$^{+}$/H$^{+}$)$_{mean}$=6.17, so that the total sulphur abundance in B can be written as S/H$\sim$(S$^{+}$/H$^{+}$)$_{mean}$ + (S$^{++}$/H$^{+}$)$_{B}$. We assumed that S$^{3+}$/H$^{+}$ is negligible. Results are presented in Table \ref{table:abinfrared} and discussed in Sect. \ref{chemical}.

\begin{table}[h!]
\caption{Lines measured over the two spectra studied in the infrared range. Integrated fluxes are in units of 10$^{-5}$~erg~cm$^{-2}$~s$^{-1}$. }
\label{table:infrared} 
\centering 
\begin{tabular}{l c c c}
\hline
&&\multicolumn{2}{c}{F($\lambda$)} \\
\cline{3-4}
Line & $\lambda$~($\mu$m) & Region A & Region B \\
\hline \hline \\
{[}S{\sc iv}] & 10.51 & ...& 4.1 $\pm$ 0.5 \\
H(7-6) & 12.37 & 4.9 $\pm$ 0.5 & 5.1 $\pm$ 1.0 \\
{[}Ne{\sc ii}] & 12.81 & 121.7 $\pm$ 3.2 &105.3 $\pm$ 3.5 \\
{[}Ne{\sc iii}] & 15.56 & 5.0 $\pm$ 0.4 & 1.1 $\pm$ 0.3 \\
{[}S{\sc iii}] & 18.71 & 133.9 $\pm$ 5.5 & 99.2 $\pm$ 1.9 \\
{[}S{\sc iii}] & 33.48 & 156.6 $\pm$ 5.3 & 135.2 $\pm$ 4.7 \\
\\
\hline
\end{tabular}
\end{table}

\begin{table}[h!]
\caption{Ionic and total chemical abundances estimated in Regions A and B with infrared spectroscopy.}
\label{table:abinfrared} 
\centering 
\begin{tabular}{l c c}
\hline
& Region A & Region B \\
\hline \hline \\
12+log(Ne$^{+}$/H$^{+})$ & 7.56 $\pm$ 0.04 & 7.47 $\pm$ 0.08 \\
12+log(Ne$^{2+}$/H$^{+})$ & 5.85 $\pm$ 0.05 & 5.18 $\pm$ 0.16 \\
12+log(S$^{2+}$(18.71$\mu$m)/H$^{+}$) & 6.59 $\pm$ 0.09 & 6.44 $\pm$ 0.17\\
12+log(S$^{2+}$(33.48$\mu$m)/H$^{+}$) & 6.64 $\pm$ 0.16 & 6.56 $\pm$ 0.21\\
12+log(S$^{3+}$/H$^{+}$) & ... & 4.41 $\pm$ 0.19 \\
\\
12+log(Ne/H) & 7.57 $\pm$ 0.04 & 7.48 $\pm$ 0.08\\
12+log(S/H)$\dagger$ & 6.72 $\pm$ 0.07 & 6.63 $\pm$ 0.11  \\
\hline
\end{tabular}
        \begin{list}{}{}
			\item {$\dagger$} Assuming S$^{+}$/H$^{+}$ from the optical spectroscopy.\\
		\end{list}

\end{table}

\section{Dicussion \label{discussion}}
We included M1-67 in our IFS observational programme to provide answers to some questions that still surround this object: degree of gas homogeneity (both kinematic and chemical), stellar evolutionary phase origin of the gas, interaction with the ISM, influence of the star spectral type at WR stage, etcetera. To do this, we put together our results for the optical (1D + 2D) and infrared analysis, and we complemented them with theoretical models of stellar evolution and previous kinematic studies of this nebula.

\subsection{Chemical content of M1-67 \label{chemical}}
The chemical abundances derived from the 1D optical and infrared studies, presented in Tables \ref{table:abinfrared} and \ref{table:paramyabun}, give us relevant information on the chemical content across the nebula. To compare the derived abundances with the expected ISM values at the location of the nebula, we use the solar values from \citet{2009ARA&A..47..481A} as our primary reference. For the sake of consistency as reference for M1-67, we consider here gas abundances derived following the same methodology, i.e. H{\sc ii} region collisional emission lines. We adopted the chemical abundances of the prototypical H{\sc ii} region M\,42 as a reference (\citealt{2007A&A...465..207S,2011MNRAS.412.1367T} with t$^{2}$=0, and references therein). Then, we corrected the t$^{2}$=0 abundances from the effect of the radial abundance gradient of the Milky Way \citep{2006ApJS..162..346R} to the galactocentric radius of M1-67\footnote{Assumed R$_{G}\sim$10 kpc as the the galactocentric distance of the representative ISM at the location of M1-67 \citep{1992A&A...259..629E} and taking the distance from the Sun to M\,42 into account (d$\sim$0.414\,kpc, \citealt{2007A&A...474..515M}).}. We considered the constant ratio $\log \mathrm{(Ne/O)}$=-0.73$\pm$0.08 since they are products of the same nucleosynthesis. After these corrections the expected ISM abundances to be compared with M1-67 are 12+$\log \mathrm{(O/H)}\sim$8.42$\pm$0.03, 12+$\log \mathrm{(N/H)}\sim$7.54$\pm$0.09, 12+$\log \mathrm{(S/H)}\sim$6.99$\pm$0.12, and 12+$\log \mathrm{(Ne/H)}\sim$7.69$\pm$0.09.\\

First of all, it can be observed that our derived oxygen abundances in R5 and R6 (12+$\log \mathrm{(O/H)}\sim$7.73$\pm$0.06, 7.67$\pm$0.07, respectively) are substantially lower than the expected value by factors $\sim$10 with respect to the solar reference, and $\sim$7 with respect to the ISM. This result implies that in the M1-67 nebula oxygen is strongly under-abundant. Comparing the derived N/H abundance with the expected ISM value, we find that nitrogen is strongly enriched in M1-67 (factor $\ge$ 6).

Overall, this chemical composition can be seen in all the nebular regions observed; the N/O ratio appears extremely high due to the effect of both nitrogen enhancement and oxygen deficiency. This fact can be understood when assuming we are seeing regions composed of material processed in the CNO cycle. This result for N/O abundance is consistent with previous 1D studies \citep{1991A&A...244..205E}, but here it has been extended across the whole (2D) nebular geometry and physical conditions. The only region where the N/H abundance is close to the ISM expected value is R7 (the region with different properties in the 2D analysis, see Sect. \ref{2d}).

We did not estimate the total helium abundances since our helium lines are very faint and the measures uncertain. Nonetheless, given the low limit of the value of He{\sc i} ($<$0.03), the absence of He{\sc ii} and the ICF inferred from \citet{2007ApJ...662...15I} (ICF(y$^{+}$)$\gg$1), we deduced that in M1-67 the largest part of helium is unseen and in neutral form. \\

The analysis of the chemical abundances obtained here is reinforced by the information derived from the infrared study. The infrared spectrum allowed us to derive the sulphur and neon abundances for the main ionic species, Ne$^{+}$, Ne$^{++}$, S$^{++}$, and S$^{+3}$. The total neon abundance derived in M1-67 is consistent within the errors with the expected ISM abundance for the two apertures (Table \ref{table:abinfrared}). The noble gas neon is not expected to suffer nucleosynthetic transformation in the stellar interior, and its abundance should be preserved.

In the case of sulphur, the derivation of the total abundance requires the contribution of the optical S$^{+}$ to be added to the ionic fractions derived from the infrared. Once this approximation has been assumed, the total abundance of S/H obtained is close to, though still slightly lower than, the expected ISM value at the galactocentric distance of M1-67. Thus we cannot rule out the possibility that the nebular material could be slightly sulphur-poor: either a certain degree of depletion on dust or maybe a nucleosynthetic origin (or both) could be at work as reported for some planetary nebulae \citep{2012ApJ...749...61H}.\\

Taking the abundance ratios into account, we can obtain clear indications of the excitation degree of the nebula. The values N$^{+}$/N~$\sim$1 and O$^{+}$/O$^{++}$~$>$1 from the optical and the derived ratios of Ne$^{+}$/Ne$^{++}$ and S$^{++}$/S$^{3+}$ from the IR study point to the very low ionization degree of the gas in M1-67. The ionization parameter obtained from the photoionization model of R5, $\log\mathrm{(U)}=-3.84$, is fully consistent with this very low excitation observed.\\

To provide a summary of the chemical abundances obtained across the nebula in the optical and infrared ranges, we have grouped regions with similar physical and chemical properties (whenever possible). In Table \ref{table:summary} we show the results: $<$1,2,3$>$ represents the average of R1, R2, and R3, $<$5,6$>$ the average of R5 and R6, and $<$A,B$>$ the average of zones A and B from the IR study. In these cases the corresponding parameters were estimated as the mean weighted by the error in each zone. The two last columns represent the expected ISM values and solar abundances from \citet{2009ARA&A..47..481A}, respectively.

\begin{table*}
		 \caption{Summary of inferred properties in M1-67.}   
		\label{table:summary}    
		\centering                      
		\begin{tabular}{l c c c c c c c}
		\hline
		& $<$1,2,3$>$ & 4  & $<$5,6$>$ & 7 & $<$A,B$>$ & ISM$^{a}$  & Solar$^{b}$ \\
		\hline
		\hline
   		\\
		c(H${\beta}$) &  1.87 $\pm$ 0.01  & 1.87 $\pm$  0.01 & 1.90 $\pm$ 0.02 & 2.15 $\pm$ 0.04 & ...  & ... & ... \\
		n$_{\mathrm{e}}$([S{\sc ii}]) (cm$^{-3}$)& 1581 $\pm$ 49 & ... & 677 $\pm$ 62 & ...  & ...  & ...  & ...\\
		12+log(O/H)  & ...  &... & 7.70 $\pm$  0.03  & ... & 8.28 $\pm$ 0.09 $^{c}$  & 8.42$\pm$ 0.03 & 8.69 $\pm$ 0.05\\
		12+log(S/H)  & 6.35 $\pm$ 0.02 & ... & 6.40 $\pm$ 0.02   & ...  & 6.69 $\pm$ 0.04 & 6.99$\pm$ 0.12  & 7.12 $\pm$ 0.03\\
		12+log(N/H) & 8.13 $\pm$ 0.01 & 8.36 $\pm$ 0.03  & 8.21 $\pm$ 0.03 & 7.92 $\pm$ 0.03 & ... & 7.54 $\pm$ 0.09 & 7.83 $\pm$ 0.05  \\
		12+log(Ne/H) & ...  & ... & ...  & ...  & 7.55 $\pm$ 0.04 & 7.69 $\pm$ 0.09  & 7.93 $\pm$ 0.10\\
		$\Delta$(log(N/H))$^{d}$ & 0.59 $\pm$ 0.09 & 0.82 $\pm$ 0.09 & 0.67 $\pm$ 0.10 & 0.38 $\pm$ 0.09 & ...  & ...& ...  \\
		$\Delta$(log(O/H))$^{d}$& ...  & ...  & -0.72 $\pm$ 0.04 & ...  & -0.14 $\pm$ 0.09 $^{c}$  & ... & ... \\
		\hline
		\end{tabular}
        \begin{list}{}{}
			\item {$^{a}$} Expected ISM abundances at R$_{G}\sim$10 kpc. \\
			\item {$^{b}$} Solar abundances from \citet{2009ARA&A..47..481A}.\\
			\item {$^{c}$} Estimated assuming $\log \mathrm{(Ne/O)}$=-0.73$\pm$0.08.\\	
			\item {$^{d}$} Variations with respect to the expected ISM abundance. \\ 
		\end{list}
		\end{table*}

\subsection{M1-67 structure \label{structure}}
Although the first observations of M1-67 showed a nearly spherical shape, the high contrast achieved by coronographic studies in the inner regions made a bipolar symmetry clearly visible \citep{1995IAUS..163...78N}. Owing to the field of view of our PPAK observations, we cannot detect this bipolarity; however, the narrow-band images from the INT and the interpolated maps from PPAK (see Figs. \ref{fig:rgb} and \ref{fig:morphology_all}) show that the bright knots are aligned along a preferred axis with \textquotedblleft holes\textquotedblright ~in the perpendicular direction. The integrated spectrum of R4, confirms that the emission in the holes is very faint (i.e. H${\beta}$ was not detected). Furthermore, the MIPS image from Spitzer (Fig. \ref{fig:spitz_24micr}) also reveals the bipolar appearance at 24$\mu$m, suggesting that the ionized gas is mixed with warm dust. We emphasize that the knots are not only regions with high surface brightness but also very dense areas where the [N{\sc ii}]/H${\alpha}$ and [N{\sc ii}]/[S{\sc ii}] ratios show the maximum values.\\

We support the idea of a preferred axis, but either way, is the bipolarity the footprint of an ejection from the star? Looking at the radial velocity map of Fig. \ref{fig:ha_velocity}, we can see that velocity decreases when we move away from the centre (except in the far NE where the gas has peculiar properties, see below). This agrees with the studies from \citet{1981ApJ...249..586C}, who predict a faster movement near the star, and \citet{1998A&A...335.1029S} who discuss the idea of a bipolar outflow. The spatial distribution of the electron density shows similar behaviour: the mean values of maps and integrated spectra of the central pointing are higher than at the edge ($\sim$1500~cm$^{-3}$ and $\sim$650~cm$^{-3}$, respectively). Also the electron density decreases along the radial cut seen in Fig. \ref{fig:dens_rad} with a symmetric gradient in two directions (from the centre to NE and to SW) and flattening towards the edges. Both analyses lead us to think that the preferred axis is not only morphological, but is also the footprint of a mechanism that could have expelled material in the past and, later, interacted with the ISM diluted and decelerated the gas.\\

\begin{figure}
\centering
\includegraphics[width=9cm]{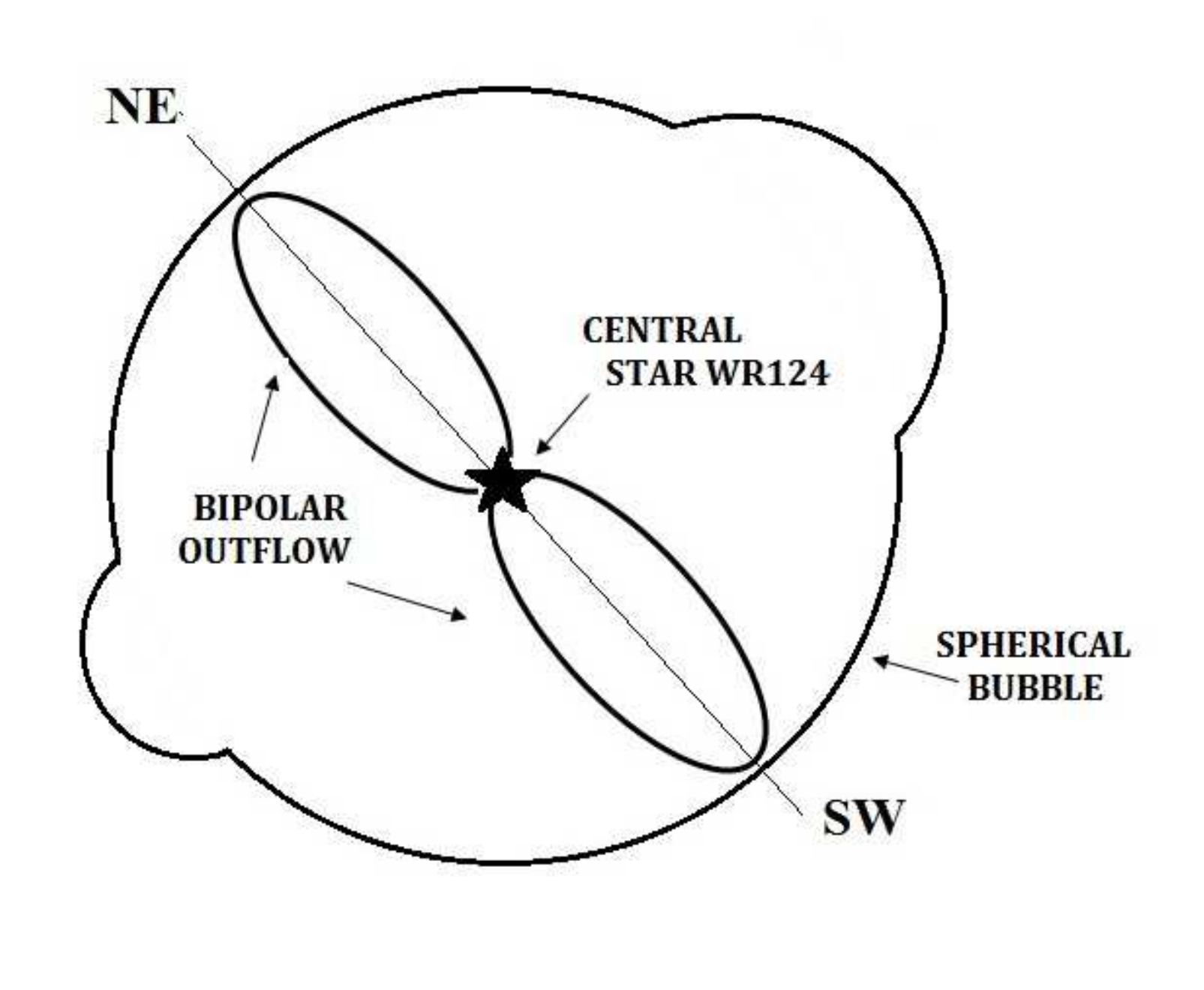}
\caption{Sketch showing the structure of M1-67 around the central star WR124: the bipolar axis along the direction NE-SW and the spherical bubble.}
\label{fig:sketch}
\end{figure}

Leaving aside for a moment the discussion of bipolarity, there is another striking morphological feature in this object. The IR study at 24$\mu$m reveals a spherical bubble surrounding the bipolar structure. Kinematic studies from \citet{1998A&A...335.1029S} show two different motions in the enviroment of WR124: a bipolar outflow and an external spherical hollow shell expanding into the ISM. In our narrow-band images from the INT, this bubble is not detected possibly because the material is diluted in the ISM and very weak in the optical range. A simple sketch of the proposed structure of M1-67 is presented in Figure \ref{fig:sketch}: an inner region with bipolar or elliptical shape along the direction NE-SW surrounded by an external spherical bubble.\\

The study of the edge pointing region suggests that the gas in the NE possesses different properties from the gas of the bipolar outflow. In summary, the properties found for this region are a) the largest reddening coefficient of the nebula with c(H${\beta})>$2.5; b) the only area where we measure [N{\sc ii}]$<$H${\alpha}$ and with the smallest N/H abundance estimated, close to the solar neighbourhood value; c) increase rather than decrease in the relative radial velocity; d) absence of [S{\sc ii}]$\lambda\lambda$6717/6731emission lines; e) the minimum H${\beta}$ flux measured (R7) and lack of [O{\sc iii}]$\lambda$5007 or helium lines. The presence of these properties puzzles us, but here we propose a possible scenario to explain the origin of this region. The nitrogen abundance of our \textquotedblleft peculiar\textquotedblright ~region points towards material not processed in the CNO cycle (e.g. ISM or MS bubble) while the morphology (Fig. \ref{fig:rgb}) and kinematics (Fig. \ref{fig:ha_velocity}) suggest that it does not belong to the bipolar ejection. When looking at the bow-shock simulations of \citet{2003A&A...398..181V} and taking the external IR bubble into account, it is possible that the high velocity of the runaway WR124 causes a paraboloid-like bow shock around the star sweeping up the surrounding medium, so that we are seeing the remaining of this bow-shock placed in our line of sight. We should bear in mind that the peculiar region is spatially close to the small reversed bow-shock like structures found by \citet{2001ApJ...562..753G} at the NE periphery of M1-67.

\subsection{M1-67: a consequence of the evolution of the central star WR124 \label{evolution}}
The theory of evolution of massive stars can help us explain the observed structure. We compare the stellar parameters from the central star in M1-67 (effective temperature and luminosity from \citealt{2006A&A...457.1015H}) with the stellar evolution models from STARS \citep{1971MNRAS.151..351E,1995MNRAS.274..964P,2004MNRAS.353...87E}, \citet{2003A&A...404..975M}, and the most recent models from \citet{2012A&A...537A.146E} to estimate the initial mass of the WR star. Regardless of small discrepancies, all the models predict initial mass for WR124 of 60-80~M$\sun$ (J. Toal\'a, \textit{private communication}). The evolutionary scenario for a single massive star with 60~M$\sun<$M$_{\mathrm{i}}<$90~M$\sun$ follows the sequence O-Of/WNL$\longleftrightarrow$LBV-WNL-WCL-SN \citep{2011BSRSL..80..266M}. After spending a normal life as O stars on the main sequence (MS), they evolve towards cooler temperatures becoming luminous blue variables (LBVs) \citep{1994PASP..106.1025H}. These stars undergo extremely strong mass loss (up to 10$^{-3\ldots -4}~$M$\sun$~yr$^{-1}$) through winds and occasionally giant eruptions, and thus peel off parts of their stellar envelope to form a small LBV nebulae (LBVN) \citep{1995ApJ...448..788N}. LBV stars lose their mass so fast that they rapidly evolve away from LBV stage to become WR stars. With an initial mass range of 60-80~M$\sun$, we can derive that the central star in M1-67 has experienced an LBV phase instead of a red or a yellow supergiant phase before becoming a WR star. This idea is in good agreement with previous studies of the nature of M1-67 based on different observational approaches: M1-67 is very likely the imprint of a previous LBV wind instead of a classical red super-giant (RSG) wind-blown nebula \citep{1998ApJ...506L.127G, 2003A&A...398..181V}.\\

The spectral type of the central star (WN8), tells us that it is a \textquotedblleft young\textquotedblright  ~Wolf-Rayet and that it has most likely entered the WR phase recently. Under this hypothesis, we propose that the WR winds have not had enough time to substantially interact with the previous nebular material and, therefore, the layers and observed features originate in stellar material ejected during the MS and/or LBV phases. Considering a representative expansion velocity and the linear size of the nebula, we estimate that the ejection happened $\sim$5$\times$10$^{4}$~yr ago. This value is slightly higher than the LBV phase duration ($\sim$1.3$\times$10$^{4}$~yr, \citealt{1996A&A...305..229G}) thus supporting the hypothesis that the star has recently entered the WR phase.\\

Taking the physical sizes and morphologies from an hydrodynamical simulations of a 60 M$\sun$ star as reference \citep{1996A&A...305..229G}, it is possible that the external bubble of M1-67 contains material expelled during the MS phase, which is very tenuous in the optical because of the dilution with the ISM. \citet{Castor1975} and \citet{Weaver1977} both built models the derive analytical solutions for the dynamic evolution of shock bubbles created by interaction between the ISM and the stellar wind in the MS phase.\\ 

Several observational reasons have led us to think that the bipolar ejection (or axis of preference) is composed by material ejected during the LBV stage. First, the abundances in the knots along this axis show enrichment in nitrogen and deficiency in oxygen, a behaviour typical of CNO-processed material in phases after the MS stage\footnote{We should bear in mind that models that include rotation \citep{Meynet2005} and recent observations of O stars in the LMC \citep{2012A&A...537A..79R} have revealed that some stars of the MS stage can also show CNO-processed material.}. It is common in observations of LBVN to find very intense [N{\sc ii}] emission and absence of [O{\sc iii}] \citep{1995ApJ...448..788N,1998ApJ...503..278S,2002A&A...393..503W}, this being also indicative of a low effective temperature and low degree of excitation. Furthermore, many of these nebulae show clumpy radial structures (not multiple shells) and morphologies with preferred axes \citep{1993ApJ...410L..35C}. The presence of a bipolar ejection in M1-67 enhances the similarity of the nebula to other LBVN, which almost all display some degree of bipolarity \citep{1995ApJ...448..788N, 2001RvMA...14..261W}. In short, M1-67 shows the general properties of LBV nebulae: linear size, total ionized gas, velocity field, IR emission, chemical abundances, line intensities, and dynamical characteristics; this clearly points to an LBV progenitor \citep[among others]{1995ApJ...448..788N, 2001ApJ...551..764L,2011BSRSL..80..440W}.

The idea of M1-67 being made up of material ejected during the LBV stage was suggested in the past by \citet{1998A&A...335.1029S} based on the total mass of ionized gas, the expansion velocity, and the linear size of the nebula. Also \citet{1998ApJ...506L.127G} explain the clumpy appearance of M1-67 by assuming the interaction of winds in a previous LBV phase.\\

Our study depicts M1-67 as a nebula with two regions: an external spherical bubble with material likely produced during the MS and an inner nearly elliptical region along the NE-SW direction produced due to an ejection in the LBV phase. We are observing a WR nebula with LBVN appearance.\\

\section{Summary and conclusions \label{conclusions}}
\renewcommand {\labelenumi} {\arabic {enumi}$)$}
\renewcommand {\labelenumii} {$\bullet$}
In this work, we have presented the first integral field spectroscopy study of the ring-nebula M1-67 around the Wolf-Rayet star WR124 in the optical range with PPAK. Two regions of the nebula were observed and analysed by means of 2D and 1D studies. We also obtained and analysed IR spectroscopic data and the MIPS 24$\mu$m image of M1-67 from Spitzer. In the following, we present the main results derived from this work.

\begin{enumerate}
\item We obtain maps from the emission lines that allow us to perform a detailed study of the 2D structure of the nebula:
\begin{enumerate}
\item Interpolated maps from the main emission lines show a clumpy structure with bright knots aligned along a preferred axis in the NE-SW direction. The [O{\sc iii}]$\lambda$5007\AA{} emission is absent over the whole nebula.
\item The spatial distribution of the reddening coefficient maps, c(H${\beta}$), presents slight variations between the two pointings. In the central region c(H${\beta}$) ranges from 1.3 to 2.5 with a mean value of $\sim$1.85, while in the edge pointing the mean is 2.11, ranging from 1.7 to 2.8.
\item Electron density maps, n$_{e}$, derived from the [S{\sc ii}]$\lambda\lambda$6717/6731 ratios, show a non-uniform structure. Knots with higher surface brightness in H${\alpha}$ possess the highest densities. We also find that density decreases with increasing the distance from the star showing a symmetric gradient.
\item We analysed the ionization structure by means of line ratios maps. In particular, the [N{\sc ii}]/H${\alpha}$ map of the edge pointing field reveals two behaviours, thus defining two spatially well delimited regions: one in the NE with [N{\sc ii}]$<$H${\alpha}$ and the second one with [N{\sc ii}]$\ge$H${\alpha}$.
\item With radial velocity maps we studied the kinematics of the nebula. The derived heliocentric velocity for M1-67 is $\sim$139~km~s$^{-1}$, in agreement with previous results. The relative radial velocity seems to decrease as it moves away from the central star along the preferred axis. \\
\end{enumerate}

\item We derived the physical parameters and chemical abundances of M1-67 using the integrated spectra of eight regions:
\begin{enumerate}
\item The electron densities inferred on the central region present higher values than on the edge ($\sim$1500~cm$^{-3}$ and  $\sim$650~cm$^{-3}$, respectively). This result agrees with the radial variations of the 2D study.
\item We derived an electron temperature of $\sim$8200~K  in R5 by using our measurement of the [N{\sc ii}]$\lambda$5755\AA{} emission line.
\item The chemical abundances show, in all the studied areas, an enrichment in nitrogen and a deficiency in oxygen. The nitrogen enhancement in each region is different, suggesting an inhomogeneous chemical enrichment.\\
\end{enumerate}

\item The 24$\mu$m image reveals an inner bipolar-like structure in the NE-SW direction and an outer faint spherical bubble interacting with the surrounding ISM. From the low-resolution mid-IR spectroscopic data, we measured the main emission lines and estimate ionic and total chemical abundances, verifying the low ionization degree of the gas. \\

\item Overall, this study revealed the clumpy structure of M1-67 with knots aligned along a preferred axis and with \textquotedblleft holes\textquotedblright~ along the perpendicular direction. The gas along this bipolar axis possesses a low ionization degree, and it is well mixed with warm dust. The optical analysis of these knots revealed chemical abundances typical  of material processed in the CNO cycle, suggesting that the material comes from an evolved stage of the star. The radial variations in electron density and velocity indicate that the gas of the bipolar feature was ejected by the star. \\

\item A region placed to the NE of the nebula shows different kinematic, chemical, and morphological properties. We propose that this region comprises the remaining of a bow-shock caused by the runaway WR124 with ISM material mixed up with the MS bubble.\\
\end{enumerate}

Based on our observational results and taking theoretical models from the literature into account (e.g. \citealt{1996A&A...305..229G}), we propose a scenario where the central star has recently entered the WR phase. This implies that the interaction of WR winds with previous surrounding material is not visible yet. After comparing our results with stellar evolution models and taking the inferred initial mass of the star (60~M$\sun < $ M$_{i} <$ 80~M$\sun$)  into account, we deduced that the central star experienced an LBV stage before becoming a WR. The bipolar material observed belongs to an ejection during the LBV stage since the morphology, kinematics, and chemistry are in good agreement with previous studies of LBV nebulae.\\


\begin{acknowledgements}
This work is supported by the Spanish Ministry of Science and Innovation (MICINN) under the grant BES-2008-008120. This work has been partially funded by the projects: AYA2010-21887-C04-01 of the Spanish PNAYA and CSD2006 - 00070 "1st Science with GTC from the CONSOLIDER 2010 programme of the Spanish MICINN and TIC114 of the Junta de Andaluc\'ia. We thank J. Toal\'a for providing estimations for the initial mass of the WR star and for useful suggestions. We are also very grateful to M. Fern\'andez-Lorenzo, A. Monreal-Ibero, K. Weis, and the ESTALLIDOS collaboration for their useful comments and scientific support.
\end{acknowledgements}



\begin{landscape}
\begin{table}
\caption{Measured lines in all the integrated spectra. The intensity values of each emission line were normalized to $F(H{\beta})=100$ and reddening-corrected.}  
\label{table:all_lines}   
		\centering                  
		\begin{tabular}{l c c c c c c c c c c }
		\hline
        &&&\multicolumn{8}{c}{I($\lambda$)/I(H${\beta}$)} \\
       \cline{4-11}
		Line & $\lambda~(\AA{})$ & $f(\lambda)$ & Region 1 & Region 2  & Region 3 & Region 4$^{a}$ & Region 5 & Region 5 (model)$^{b}$ & Region 6 & Region 7  \\
		\hline
		\hline         \\
		{[}O{\sc ii}]   & 3727 & 0.322 &  ... &  ...  &  ...  & ...  & 47.5 $\pm$ 4.0 $\dagger$ &  50.2 &...  &  ...    \\
		H7 & 3970 & 0.266 & ...  &  ...  & 26.3 $\pm$ 1.8 $\dagger$&  ...  &  ...  & 15.8 & 18.2 $\pm$ 1.5 $\dagger$ &  ..  \\
		H${\delta}$   & 4102 & 0.229 & 34.0 $\pm$ 4.8 $\dagger$& 27.0 $\pm$ 2.2 $\dagger$&  ... & ...  & 27.7 $\pm$ 1.9 $\dagger$&25.7&  33.9 $\pm$ 2.3 $\dagger$&  ...    \\
		H${\gamma}$  & 4340 & 0.156 & 50.0 $\pm$ 2.2 & 47.7 $\pm$ 2.5 & 44.4 $\pm$ 1.6 & ...  & 51.4 $\pm$ 2.5 & 46.6 & 52.2 $\pm$ 2.6 & 64.9 $\pm$ 1.5 $\dagger$\\
		H${\beta}$  & 4861 & 0.000 & 100.0 $\pm$ 0.7 & 100.0 $\pm$ 0.8 & 100.0 $\pm$ 0.5 & 100.0 $\pm$ 3.2 & 100.0 $\pm$ 0.3 &100.0&  100.0 $\pm$ 0.9 & 100.0 $\pm$ 1.8 \\
		{[}O{\sc iii}]  & 5007 & -0.038 & ...  & ...  &  ...  &  ...  & 1.8 $\pm$ 0.1 $\dagger$& 2.0 &  0.9 $\pm$ 0.1 $\dagger$&  ... \\
		He{\sc i} & 5016 & -0.040 & ...&  ...  &  ...  &  ...  & 1.2 $\pm$ 0.6 $\dagger$ & 0.5 &  ...  &  ...    \\
		{[}N{\sc ii}]  & 5755 & -0.185 &  ...  &  ...  &  ...  &  ...  & 3.2 $\pm$ 0.2 & 2.9 &1.7 $\pm$ 0.2 $\dagger$&  ...  \\
		He{\sc i}  & 5876 & -0.203 &  ...  &  ...  &  ...  &  ...  & 2.5 $\pm$ 0.1 $\dagger$& 2.5&  1.5 $\pm$ 0.1 $\dagger$&  ...   \\
       {[}N{\sc ii}]  & 6548 & -0.296 & 106.8 $\pm$ 8.2 & 104.1 $\pm$ 2.6 & 103.3 $\pm$ 4.0 & 195.5 $\pm$ 7.1 & 139.2 $\pm$ 10.5 & 173.2 & 116.6 $\pm$ 10.3 & 61.3 $\pm$ 1.9  \\
		H${\alpha}$  & 6563 & -0.298 & 324.9 $\pm$ 19.6 & 312.7 $\pm$ 7.8 & 316.2 $\pm$ 12.2 & 303.3 $\pm$ 5.1 & 331.8 $\pm$ 25.2 & 293.8 & 336.8 $\pm$ 30.0 & 304.2 $\pm$ 9.2  \\
		{[}N{\sc ii}]   & 6583 & -0.300 & 330.7 $\pm$ 21.0 & 332.4 $\pm$ 8.3 & 328.2 $\pm$ 12.8 & 538.4 $\pm$ 4.4 & 425.7 $\pm$ 32.6 & 511.1 &367.7 $\pm$ 33.1 & 202.0 $\pm$ 6.2 \\
		He{\sc i}  & 6678 & -0.313 &  ...  &  ...  &  ...  &  ...  &  ...  & 0.7& 0.3 $\pm$ 0.1 $\dagger$&  ...    \\
		{[}S{\sc ii}]  & 6716 & -0.318 & 12.6 $\pm$ 1.1 & 13.4 $\pm$ 0.4 & 12.4 $\pm$ 0.5 &  ...  & 18.9 $\pm$ 1.5 &  20.0 &16.6 $\pm$ 1.6 &  ...  \\
		{[}S{\sc ii}]   & 6731 & -0.320 & 16.8 $\pm$ 1.3 & 18.4 $\pm$ 0.5 & 16.6 $\pm$ 0.7 &  ...  & 19.7 $\pm$ 1.6 & 21.1 &18.2 $\pm$ 1.7 &  ...   \\
				\\
		F(H${\beta}$)$^{c}$  &  &  & 151.77 $\pm$ 1.10 & 168.60 $\pm$ 1.30 & 92.41 $\pm$ 0.46 & 3.21 $\pm$ 0.10 & 22.97 $\pm$ 0.08 &  ...  & 52.13 $\pm$ 0.48 & 1.89 $\pm$ 0.03  \\
		c(H${\beta}$)   &  &  & 1.85 $\pm$ 0.08 & 1.88 $\pm$ 0.04 & 1.86 $\pm$ 0.06 & 1.87 $\pm$ 0.01  & 1.88 $\pm$ 0.11 &  ...  & 1.93 $\pm$ 0.13 & 2.15 $\pm$ 0.04 \\
         \\
		\hline
		\end{tabular}
        \begin{list}{}{}
			\item {$^{a}$} In Region 4,  the H${\beta}$ flux was not measured. We estimated it by performing the inverse process of the extinction correction. See text for details.\\
			\item {$^{b}$} Emission lines intensities from  the photoionization model performed for Region 5. \\ 
			\item {$^{c}$} H${\beta}$ fluxes in units of $x10^{-16}~erg~cm^{-2}~s^{-1}$ (not corrected for extinction).\\
			\item {$\dagger$} Lines with measures uncertain. \\
		\end{list}
		\end{table}
\end{landscape}

\begin{landscape}
\begin{table}
		 \caption{Electron densities (cm$^{-3}$) , temperatures ($\times 10^{-4}$~K), ionic abundances, and total chemical abundances for the integrated spectra. }  
		\label{table:paramyabun}     
		\centering                       
		\begin{tabular}{l c c c c c c c c}
		\hline
		& Region 1 & Region 2  & Region 3 & Region 4 & Region 5 & Region 5 (model)$^{*}$ & Region 6 & Region 7  \\
		\hline
		\hline
   		\\
		n$_{\mathrm{e}}$([S{\sc ii}]) & 1484 $\pm$ 587 & 1640 $\pm$ 226 & 1504$ \pm$ 298 & 600 $\pm$ 200 $^{a}$ & 631 $\pm$ 269 & 700&  760 $\pm$ 364 & 600$\pm$ 200  $^{a}$ \\
		T$_{\mathrm{e}}$([N{\sc ii}])$\dagger$ &  ...  & ...   &  ...  & ... &  8203 $\pm$ 169 &8550 &   ... & ...  \\
		\\
12+log(O$^{+}$/H$^{+}$)  & 7.59 $\pm$ 0.06$_{E}$ & 7.63 $\pm$ 0.03$_{E}$ & 7.58 $\pm$ 0.04$_{E}$ & ... & 7.72 $\pm$ 0.06 & 7.50 &  7.67 $\pm$ 0.07$_{E}$ & ...  \\
12+log(O$^{2+}$/H$^{+}$)& ...  & ... & ... & ... & 6.10 $\pm$ 0.04 & 6.01 & 5.80 $\pm$ 0.06  & ... \\
12+log(S$^{+}$/H$^{+}$) & 6.14 $\pm$ 0.04 & 6.18 $\pm$ 0.03 & 6.13 $\pm$ 0.03 & ... & 6.22 $\pm$ 0.04 & 6.09 & 6.18 $\pm$ 0.04 & ... \\
12+log(N$^{+}$/H$^{+}$)  & 8.05 $\pm$ 0.04 & 8.05 $\pm$ 0.03 & 8.05 $\pm$ 0.03 & 8.28 $\pm$ 0.03& 8.17 $\pm$ 0.04 & 8.06 & 8.10 $\pm$ 0.04  & 7.83 $\pm$ 0.03\\
(He$^{+}\lambda$5875/H$^{+}$) & ... & ... & ...& ... & 0.020 $\pm$ 0.001 & 0.01 & 0.012 $\pm$ 0.001  & ...  \\
\\
12+log(O/H)  & ...  & ... & ... & ... &7.73 $\pm$ 0.06 & 7.62& 7.67 $\pm$ 0.07$_{E}$ & ... \\
12+log(S/H)$_{M}$  & 6.34 $\pm$ 0.04 & 6.38 $\pm$ 0.03 & 6.33 $\pm$ 0.03 & ... &6.42 $\pm$ 0.04 & 6.29 & 6.38 $\pm$ 0.04 & ... \\
12+log(N/H)$_{M}$  & 8.14 $\pm$0.04 & 8.14 $\pm$ 0.03 & 8.13 $\pm$0.03  &  8.36 $\pm$ 0.03 & 8.25 $\pm$ 0.04 & 8.14 &  8.18 $\pm$ 0.04 &  7.92 $\pm$ 0.03 \\
log(N/O)  & ... & ... & ... & ... &0.45 $\pm$ 0.07 & 0.42 &  ... & ... \\
log(N/O) N2S2 $_{E}$   & 0.46 $\pm$ 0.05 & 0.42 $\pm$ 0.02 & 0.47 $\pm$ 0.03 & ... & 0.45 $\pm$ 0.05 & ...& 0.43 $\pm$ 0.06 &  ...\\
		\\	
		\hline
		\end{tabular}
        \begin{list}{}{}
			\item {$^{*}$} Chemical abundances from  the photoionization model performed over Region 5. \\ 
			\item {$^{a}$} Assumed electron densities.\\
			\item {$\dagger$} For all the regions we used T$_{\mathrm{e}}$([N{\sc ii}]) from Region 5.\\
			\item {$_{E}$} Estimated from the empirical parameter N2S2 proposed by \citet{2009MNRAS.398..949P}. \\
			\item {$_{M}$} Total abundances estimated by using the ICFs from model of Region 5.  The sulphur values should be taken as lower limits (see Sect. \ref{chemical}).\\
		\end{list}
		\end{table}
\end{landscape}

\end{document}